\documentclass[11pt]{article}

\usepackage{amsmath,amssymb,bm, graphics,epsfig,subfigure,color}
\usepackage{amsthm,bbm,algorithm,algorithmic}
\usepackage{moreverb,booktabs,multirow,lscape}

\usepackage{url}
\usepackage{alphalph}
\usepackage{diagbox}
\newtheorem*{Lem*}{Lemma}

\numberwithin{equation}{section} \numberwithin{The}{section}

\usepackage[perpage]{footmisc}

\newcommand{\ignore}[1]{}

\begin{document}
\title{Statistical Modeling of Data Breach Risks: Time to Identification and Notification}

 \author{ Maochao Xu\footnote{Corresponding author. Email: mxu2@ilstu.edu}\qquad  Quynh Nhu   Nguyen\\
 Department of Mathematics\\
 Illinois State University, USA\\
 \date{\today}
 }
\maketitle

\begin{abstract}

It is very challenging to predict the cost of a cyber incident owing to the complex nature of cyber risk. However, it is inevitable for insurance companies who offer cyber insurance policies.  The time to identifying an incident and the time to noticing the affected individuals are two important components in determining the cost of a cyber incident. In this work, we initialize the study on those two metrics via statistical modeling approaches. Particularly, we propose a novel approach to imputing the missing data, and further develop a dependence model to capture the complex pattern exhibited by those two metrics. The empirical study shows that the proposed approach has a satisfactory predictive performance and is superior to
other commonly used models.


\end{abstract}



\section{Introduction and motivation}
Data breach is one of the most devastating risks  to computer systems, and it has become and  will continue to be a big problem due to  enormous network activities. This has been witnessed by many severe cybersecurity incidents.  For example, the Privacy Rights Clearinghouse (PRC)  reports that more than 11 billion records have been breached since 2005\footnote{https://privacyrights.org}. The Identity Theft Resource Center and Cyber Scout reports 1,108 data breach incidents in 2020, affecting 310,116,907 individuals, while the number of incidents increases to 1,862 in 2021 affecting 293,927,708 individuals\footnote{https://notified.idtheftcenter.org/s/2021-data-breach-report}. The cost of data breach is also substantial.  According to the Cost of a Data Breach Report 2021 published by IBM\footnote{https://www.ibm.com/security/data-breach}, the average cost of a data breach incident increases from
\$3.86 million in 2020 to \$4.24 million in 2021, and the average  per record cost of a data breach increases 10.3\% from 2020 to 2021 (i.e. \$146 vs \$161).

Due to the unique nature of cyber risk, the breach is often discovered several days, months, or even years later. The longer a breach goes unaddressed, the more data gets leaked and the larger the overall impact – financial and otherwise. For example, according to the 2021 IBM report,  the mean times to identifying a data breach from 2019 to 2021 are 206, 207, and 212 days, respectively, and the mean times to containing a data breach are 279, 280, and 287 days, respectively. The report also points out that the faster a data breach can be identified and contained, the lower the cost. For example, a breach with a lifecycle over 200 days costs an average of \$4.87 million while it costs \$3.61 million with a lifecycle of less than 200 days in 2021.
This discovery coincides with the simulation study in \cite{hua2020pricing} that the time to identification is the key to reduce the cost. The other important factor related to the cost of a data breach is the time to notification.  The notification allows the affected individuals to take proactive actions (e.g., changing their passwords,  and monitoring their credit scores, etc.), and hence reduces the number of potential lawsuits against the organization. However, to the best of our knowledge,  there is no formal statistical approach for modeling those two metrics: time to identification (TTI), and time to notification (TTN). The only relevant work is in  Bisogni et al. \cite{bisogni2016estimating}, where they employ the negative binomial regression to study the relationship between TTI  and different sectors.

Since the data breach has become the most common and dangerous cyber risk nowadays, there exist several studies on the statistical modeling of data breaches in the literature which are loosely related to our current study.
 For example, {Romanosky et al. \cite{romanosky2011data} 
used a fixed effect model to estimate the impact of data breach disclosure policy on the   frequency  of identity thefts incurred by data breaches.} Buckman et al. \cite{buckman2017organizations} studied the time intervals between data breaches for the enterprises that have at least two incidents between 2010 and 2016.
They showed that the duration between two data breaches may increase or decrease, depending on certain factors.
Edwards et al. \cite{edwards2016hype}
analyzed the temporal trend of data breach size and frequency and showed that the breach size follows a log-normal distribution and the frequency follows a negative binomial distribution. They further showed that the frequency of large breaches (over 500,000 breached records) follows a Poisson distribution, rather than a negative binomial distribution, and that the size of large breaches still follows a log-normal distribution.
Eling and Loperfido \cite{eling2017data} studied data breaches from the perspective of actuarial modeling and pricing. They used the  multidimensional scaling and goodness-of-fit tests to analyze the distribution of data breaches. They showed that different types of data breaches should be analyzed separately and that breach sizes can be modeled by a skew-normal distribution.  
Sun et al. \cite{sun2021modeling} developed a frequency-severity actuarial model of aggregated enterprise-level breach data to promote ratemaking and underwriting in insurance.
{Ikegami and Kikuchi \cite{ikegami2020modeling} studied a breach dataset in Japan and developed a probabilistic model for estimating the data breach risk.
They showed that the inter-arrival times of data breaches (for those enterprises with multiple breaches) follow a negative binomial distribution.} Xu and Zhang \cite{xu2021data} showed that the nonstationary extreme value model can capture the statistical pattern of the monthly maximum of data breach size very well, and they also discovered a positive time trend based on the PRC dataset. Using the same dataset, Jung \cite{jung2021extreme} compared the estimates of extreme value distributions  after 2014 and before 2014, and concluded that there is a significant increase with a break in the loss severity.  One may also refer to the recent reviews on cyber risk modelings \cite{eling2020cyber,zeller2021comprehensive,woods2021sok}.


The current study is different from those in the literature since we aim to study the statistical properties of TTI and TTN. We summarize our contributions in the following: i) Since there exist missing data for both TTN and TTI, and particularly, the missing percentage for TTI is very high, we introduce a novel copula approach to tackling this issue. Compared with  commonly used missing data imputation approaches such as Kalman Smoothing and MICE imputations, the proposed copula approach is simple but efficient, and leads to a better predictive performance. ii) A dependence model is developed to capture the positive dependence between TTN and TTI.  The empirical study shows that the proposed model is superior to other commonly used multivariate time series models. iii) The practical implications from the model results are discussed.


The rest of the paper is organized as follows. In Section \ref{sec:eda}, we conduct exploratory data analysis on the breach data to motivate the proposed model. Section \ref{sec:prel} introduces some preliminaries for the statistical modeling. In Section \ref{sec:missing}, the copula approach for imputing the missing data is introduced. In Section \ref{sec:model}, we develop the dependence model for TTN and TTI, and assess the model performance.  In Section \ref{sec:discussion}, we conclude the current study and present some discussion.


\section{Exploratory data analysis}\label{sec:eda}
The breach notification data was manually collected from the California Attorney General website\footnote{California law requires a business or state agency to notify any California resident whose unencrypted personal information is breached. \url{https://oag.ca.gov/privacy/databreach/list}}  which provides a list of breach notification reports. The report includes organization name, date(s) of breach if known,  reported date, and a brief description of incident(s).  {Since the earliest
report is on January 20, 2012, we study the time period from this earliest date to December 31, 2020 with a total of $n=2123$ breach reports.} There exist a few notifications with multiple breach dates due to the same
incident. For those notifications, the earliest breach date is used in our analysis. For instance, Steel Partners Holdings L.P. submitted a breach notification on November 23, 2020  with two breach dates April 18, 2020 and April 29, 2020, and we use the first date as the breach date since they were related to the same incident.

\begin{table}[htbp!]\centering
\begin{center}
\begin{tabular}{c|ccccccc|cc}
  \toprule
    &Min &$Q_1$ &Median &Mean &SD &$Q_3$ &Max & NAs\% &0s\%
   \\\midrule
TTN   &0.00  &47 & 102 & 189.8  &268.29 & 223 & 3222  &11.35 &0.09   \\ \midrule
TTI   & 0.00 & 0.00 & 20  &101.8  &231.81 & 96  &3140  &36.65 &18.23  \\\midrule
ITN   & 0.00 & 25  &44  &62.01  &60.92  &76  &539  &30.33 &0.71    \\ \midrule
\bottomrule
\end{tabular}
\end{center}
\caption{Statistics of TTI (time to identification), TTN (time to notification), and ITN (identification to notification)  where `SD' stands for standard deviation, $Q_1$, $Q_3$ represent the first and third quantiles and NAs for missing values.\label{table:summary}}
\end{table}

\begin{figure}[!hbtp]
\centering
\subfigure[TTN]{\includegraphics[width=.4\textwidth]{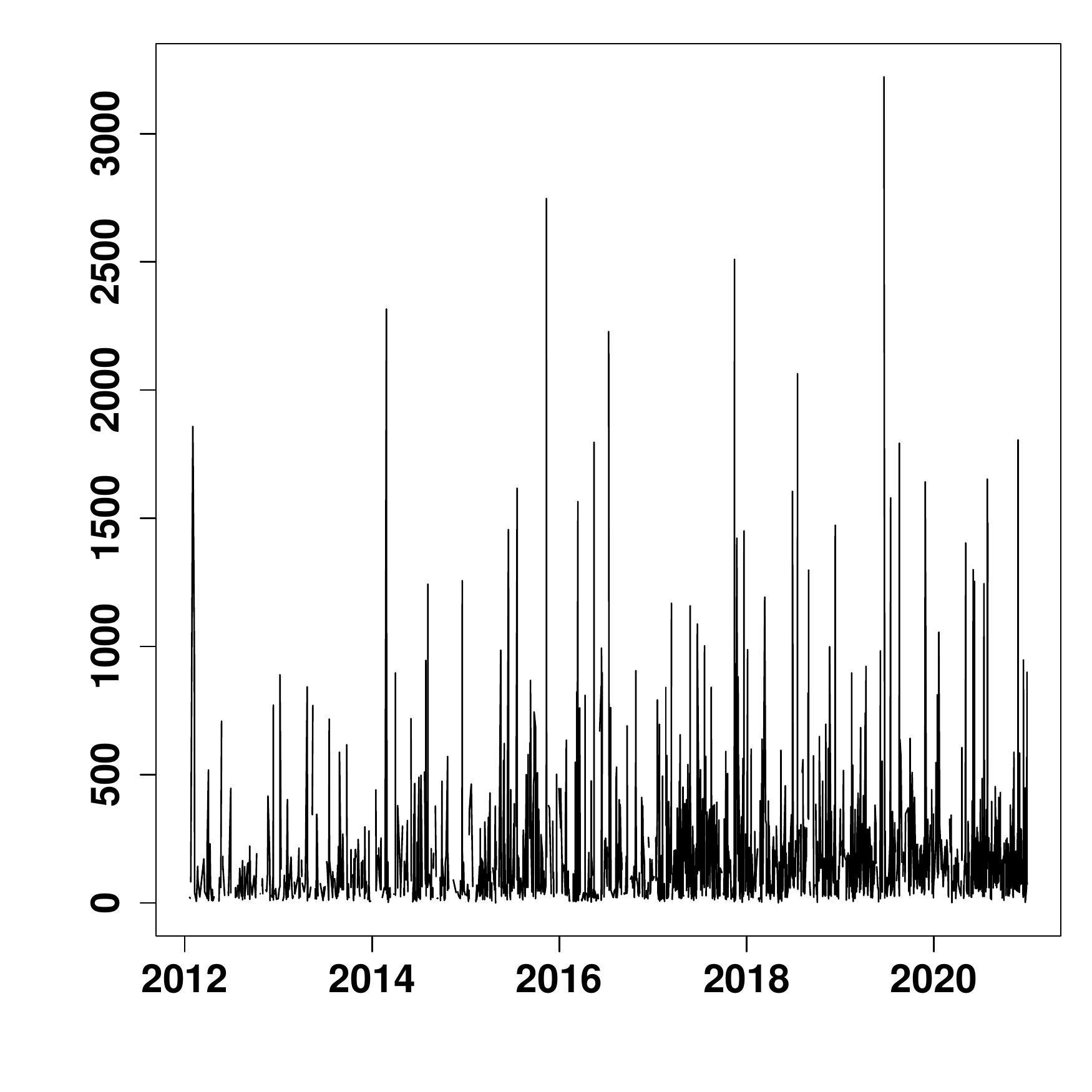} \label{fig:ts-ttn}}
\subfigure[TTI]{\includegraphics[width=.4\textwidth]{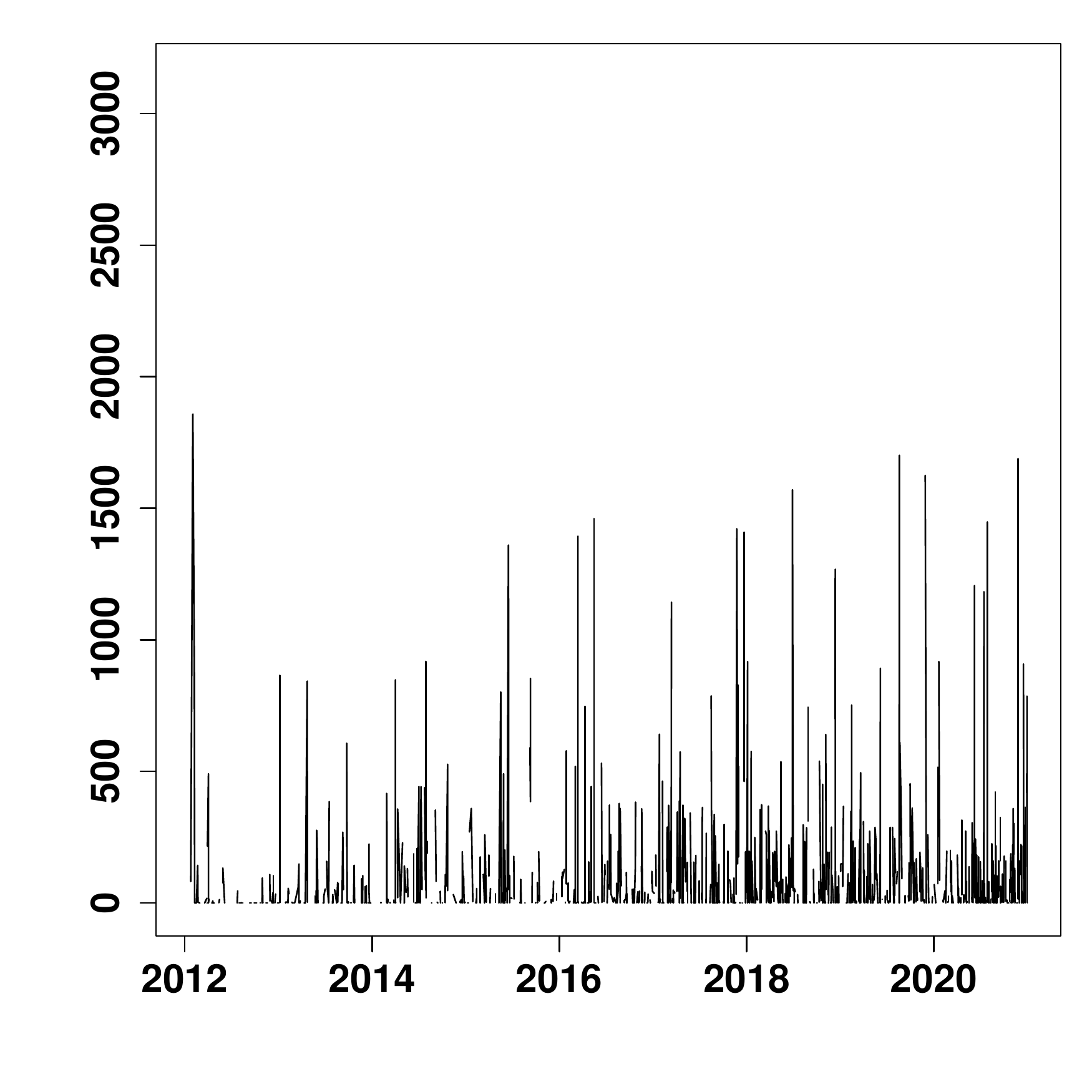} \label{fig:ts-tti}}
\caption{Time series plots of TTI and TTN,  Unit: days. \label{fig:ttplot}}
\end{figure}

\begin{figure}[!hbtp]
\centering
\subfigure[TTN]{\includegraphics[width=.32\textwidth]{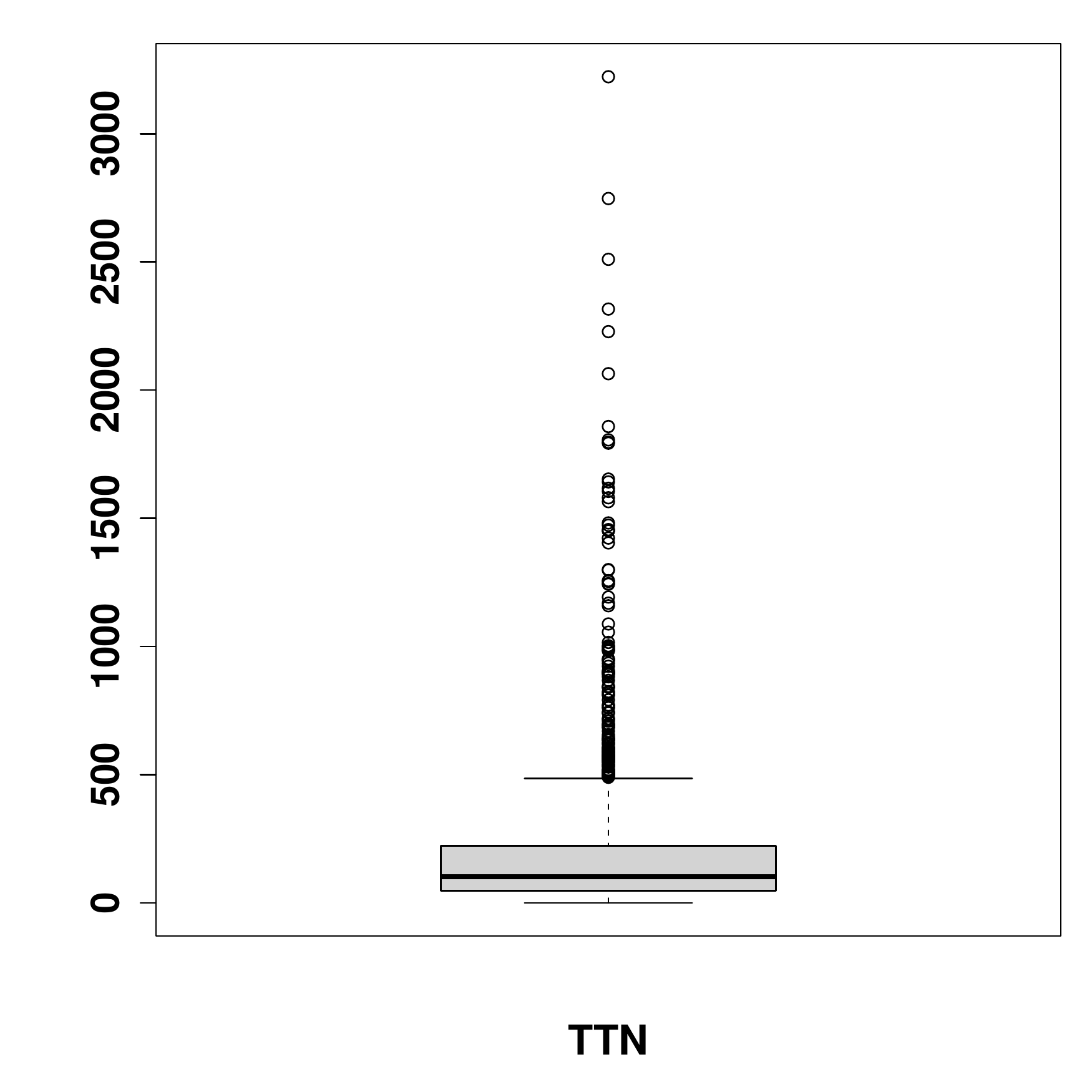} \label{fig:boxplot-ttn}}
\subfigure[TTI]{\includegraphics[width=.32\textwidth]{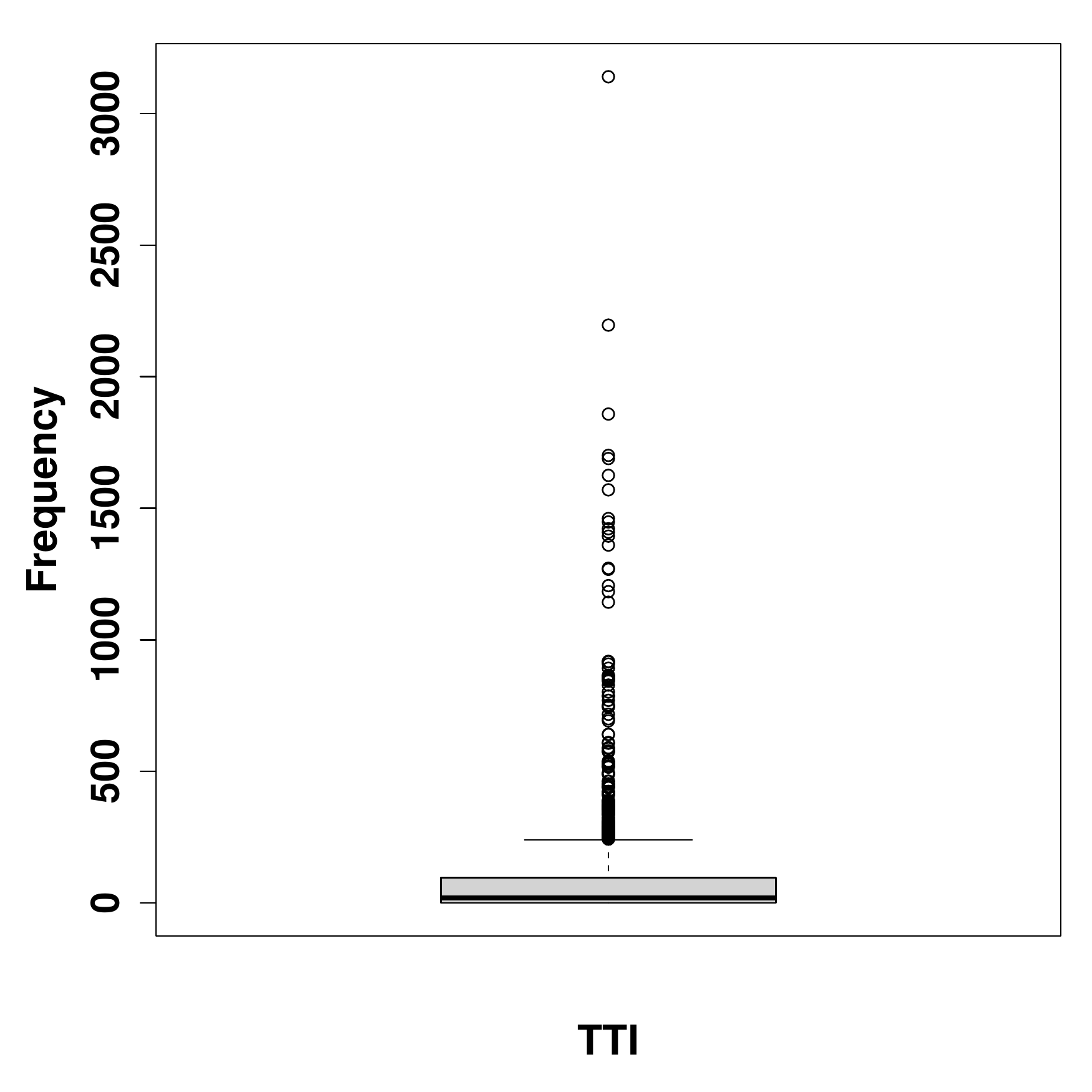} \label{fig:boxplot-tti}}
\subfigure[ITN]{\includegraphics[width=.32\textwidth]{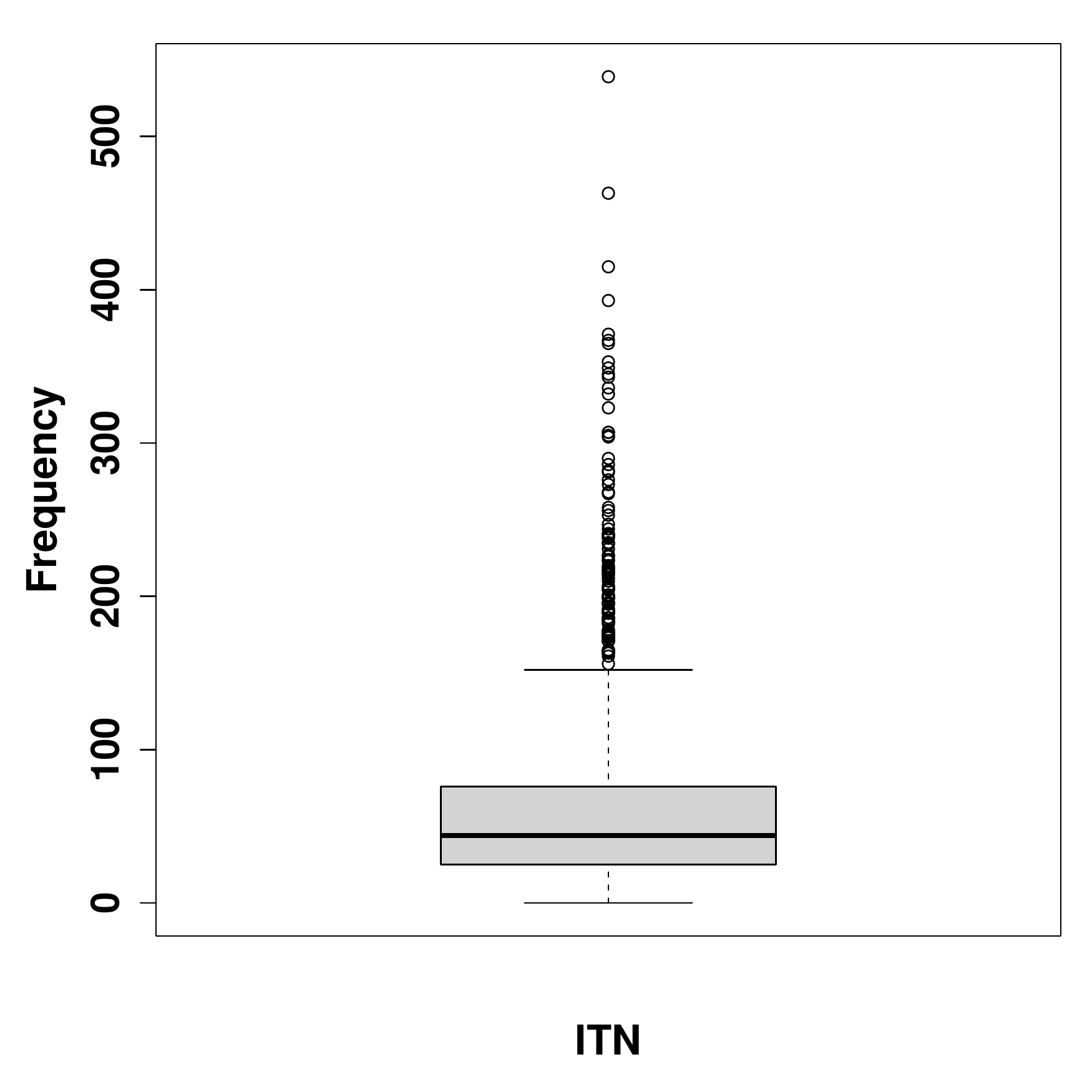} \label{fig:boxplot-itn}}
\caption{Boxplots of TTI and TTN,  Unit: days. \label{fig:ttplot}}
\end{figure}

The summary statistics of TTN is reported in Table \ref{table:summary}. It is seen that the minimum of TTN is 0 which means that the breach is reported on the same day of its occurrence. However, this percentage is small, only $0.09\%$. The mean value of TTN is 189.8 days,  and the median is $102$ days, with standard deviation $268.29$ days. The time series plot of TTN is shown in Figure \ref{fig:ts-ttn}, and we observe that there are some very large values in   TTN. This indicates that  TTN has a large variability. The boxplot in Figure \ref{fig:boxplot-ttn} shows that TTN is very skewed with a large variability.  The largest value of TTN is 3222, and it corresponds to the incident of Dominion National reported on 6/21/2019. The breach occurred as  early as 2010\footnote{https://www.newsbreak.com/news/2301704159983/dominion-national-reaches-2m-settlement-over-nine-year-data-breach}, and was
the second-largest breach reported to the Department of Health and Human Services. This incident affected 2.9 million patients and caused Dominion National a 2 million settlement. Further, there exist  $11.35\%$ missing data (i.e., unknown breach dates) for the whole time period.  

For TTI, the time series plot in Figure \ref{fig:ts-tti} shows there are some very large values and also small values. This indicates that there exists a large variability in TTI. It is seen from Table \ref{table:summary} that there are $18.23\%$ 0s, which indicates that  a small percentage of incidents can be detected on the same day of their occurrences. The mean of TTI is 101.8 days, while the median, 20 days, is much smaller. This suggests that   TTI is very skewed as seen from the boxplot in Figure \ref{fig:boxplot-tti}.  The percentage of missing data is very high, i.e., $36.65\%$. The missing data represents that the breach report does not contain either the breach date or the identification date. The largest value of 3140 days corresponds to the same incident as that of TTN.

The other quantity of interest is the time length from the identification to the notification (ITN) which is the difference of TTN and TTI.  It is seen from Table \ref{table:summary} that the mean of ITN is 62.01 days with standard deviation 60.92 days. The median is 44 days which is smaller than the mean. California data breach laws require the most expedient time possible and without unreasonable delay for the notification\footnote{see Cal. Civ. Code §§ 1798.81.5, 1798.82}. It should be noted that the data breach laws are enacted at the state level, and some states require that the notification must be made within a certain time period, ranging from 24 hours to 90 days\footnote{https://www.itgovernanceusa.com/data-breach-notification-laws}. The boxplot in Figure \ref{fig:boxplot-itn} shows that there exist many large values, say, greater than 90 days. {The maximum value of ITN is 539 days which corresponds to the incident reported on 6/14/2019 from UC San Diego Health. This incident was that participants' sensitive materials in an HIV research study were made accessible to everyone working at  Christie’s Place, a San Diego nonprofit supporting women with HIV and AIDS. The organization was criticized for being delayed to notify women affected from the breach\footnote{https://inewsource.org/2019/05/14/ucsd-data-breach-hiv-women-study/}.}

\begin{figure}[!hbtp]
\centering
\subfigure[TTN]{\includegraphics[width=.4\textwidth]{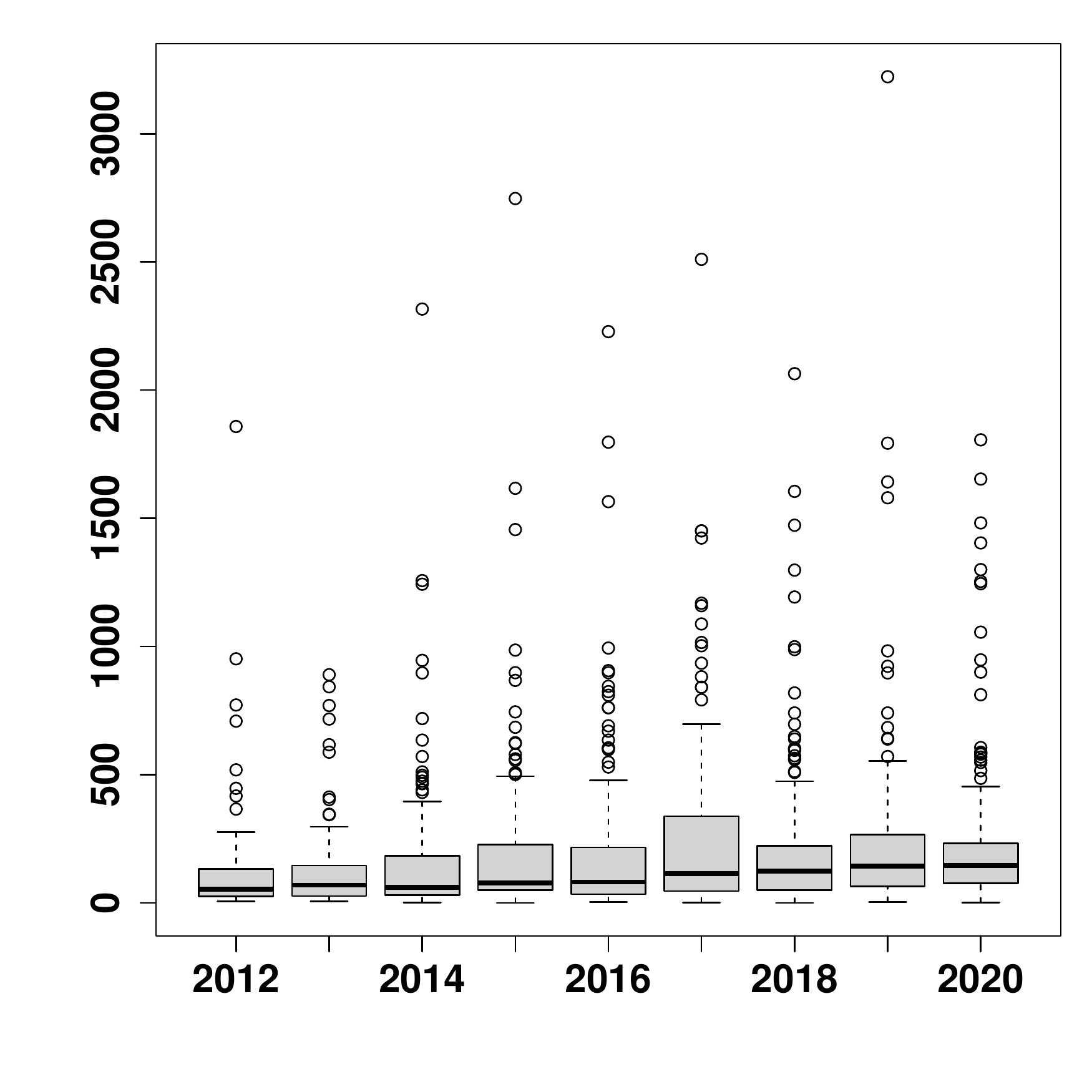} \label{fig:bpy-ttn}}
\subfigure[TTI]{\includegraphics[width=.4\textwidth]{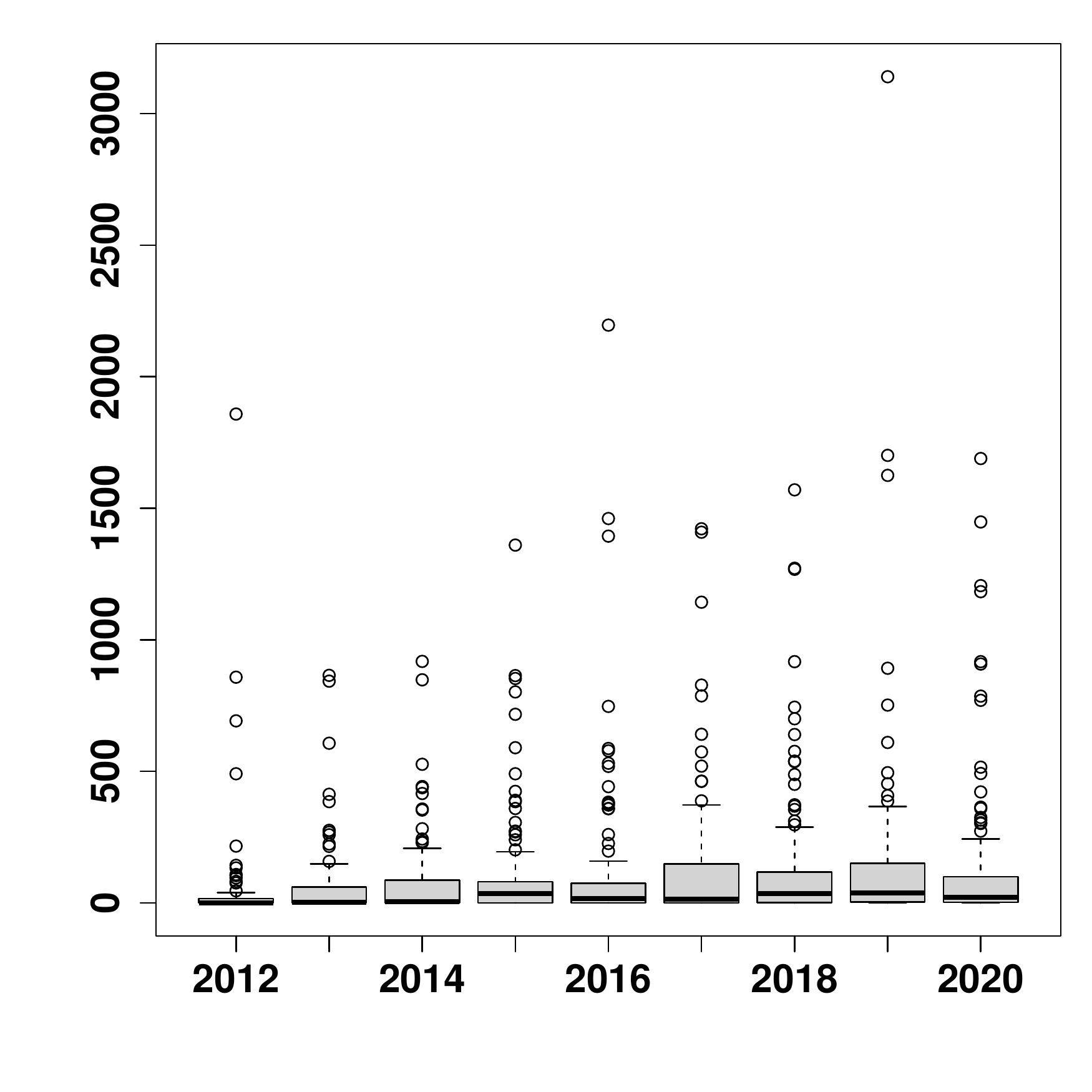} \label{fig:bpy-tti}}
\caption{Boxplots of yearly TTN and TTI \label{fig:bpyplot}}
\end{figure}

It is also of interest to see the yearly statistics of TTN and TTI. Table 
\ref{table:summary-year} shows the summary statistics of TTN and TTI. For TTN, we observe that the means in the first two years are smaller than those of other years, and the mean ranges from 122 days in 2012 to 223.19 days in 2017. The median of TTN shows an overall increasing trend. The standard deviations of TTN are very large for all years. This is further confirmed by the boxplot of TTN in Figure \ref{fig:bpy-ttn} where it shows that there exist large variabilities. The boxplot of TTN also indicates that the distribution of TTN is   heavily skewed.
\begin{table}[htbp!]\centering
\begin{center}
\begin{tabular}{c|cccccccc}
  \toprule
   Year &Min &$Q_1$ &Median &Mean &SD &$Q_3$ &Max  
   \\\midrule
 &\multicolumn{7}{c}{TTN}\\ \midrule
  2012 & 6 & 26.00 & 54.50 & 122.00 & 231.21  & 130.50 & 1858   \\ \midrule
  2013 & 6& 27.25 & 69.50 & 122.42 & 162.45 & 145.25  & 890 \\ \midrule
  2014 & 2& 31.25 & 60.00 & 165.62 & 279.60 & 181.50  & 2316 \\ \midrule
  2015 & 0& 50.00 & 78.00   & 193.20 & 299.91 & 224.20  & 2747 \\ \midrule
  2016 & 3& 34.00 & 81.50 & 183.40 & 282.95 & 212.80  & 2228 \\ \midrule
 2017 & 2& 46.25 & 114.50 & 223.19 & 287.97 & 338.00  & 2510 \\ \midrule
  2018 & 0& 51.00 & 124.50 & 196.40 & 261.50 & 222.50  & 2064 \\ \midrule
  2019 & 4& 64.25 & 143.50 & 215.20 & 308.53 & 264.50  & 3222 \\ \midrule
  2020 & 1& 77.00  & 146.00 & 199.30 & 229.37 & 232.50  & 1806 \\ \midrule
   &\multicolumn{7}{c}{TTI}\\ \midrule
  2012 & 0 & 0 & 0 & 68.72 & 251.79  & 16.50 & 1858   \\ \midrule
  2013 & 0 & 0 & 2 & 64.22 & 151.93  &  60.00  & 865   \\ \midrule
   2014 & 0 & 0 & 5.5 & 84.93 & 166.42  &  85.75  & 918   \\ \midrule
  2015 & 0 & 0.5 & 35.0 & 101.2 & 202.53  &  80.5  & 1360   \\ \midrule
  2016 & 0 & 0 & 16 & 99.78 & 263.07  &  75.00  & 2196   \\ \midrule
  2017 & 0 & 0 & 14 & 116.3 & 221.51  &  149.00  & 1422   \\ \midrule
  2018 & 0 & 1 & 36  & 114.7 &223.51  &  118.00  &1570   \\ \midrule
  2019 & 0 & 3 & 37.5  & 132.5  &313.50  &   149.5  &3140  \\ \midrule
  2020 & 0 & 2 & 21  &  93.46  &210.09  &   99.25  &1689  \\ \midrule
\bottomrule
\end{tabular}
\end{center}
\caption{Yearly summary statistics of TTI (time to identification), TTN (time to notification), and ITN (identification to notification) where `SD' stands for standard deviation, $Q_1$, $Q_3$ represent the first and third quantiles.\label{table:summary-year}}
\end{table}

 {For TTI,   the means and medians do not show any clear patterns.  There exist much more 0s in each year compared to that of TTN. The large variabities are also observed for TTI. The medians are much smaller than the means, and the boxplot of TTI in Figure \ref{fig:bpy-tti} also shows that the distribution of TTI is heavily skewed. 
 
}
 
To summarize, both TTN and TTI have considerable amounts of missing data. Particularly, the percentage  of missing data of TTI is over 36\%. Both TTN and TTI are heavily skewed and  have large variabilities. This indicates that using the mean as the risk measure of TTN/TTI is unreliable. Those properties are taken into account in our modeling process.
\section{Preliminaries} \label{sec:prel}
In this section, we introduce some preliminaries pertinent to the sequel  discussion.
\subsection{Copula}
{\em Copula}  is an effective and popular tool for modeling high-dimensional dependence and has been widely employed in many areas \cite{joe2014dependence}. Let $X_1,\ldots,X_d$ be continuous random variables with univariate marginal distributions $F_1,\ldots,F_d$, respectively. Denote their joint cumulative distribution function (CDF) by
$$F(x_1,\ldots,x_d)=P(X_1\le x_1,\ldots,X_d\le x_d).$$
A $d$-dimensional copula, denoted by $C$, is a CDF with uniform marginals in $[0,1]$, namely the joint CDF of the random vector $(F_1(X_1),\ldots,F_d(X_d))$.
Sklar's theorem \cite{joe2014dependence} says that when the $F_i$'s are continuous, $C$ is unique and satisfies
$$F(x_1,\ldots,x_d)=C(F_1(x_1),\ldots,F_d(x_d)).$$
Let $c(u_1,\ldots,u_n)$ be the $d$-dimensional copula density function and $f_i$ be the marginal density function of $X_i$ for $i=1,\ldots, d$. The joint density function of $(X_1,\ldots, X_d)$ is
$$f(x_1,\ldots,x_d)=c(F_1(x_1),\ldots,F_d(x_d))\prod_{i=1}^d f_i(x_i).$$

To model the dependence between TTN and TTI, a bivariate copula is used. Particularly, we recall the following two copulas \cite{joe2014dependence,schepsmeier2015package}:
\begin{itemize}
    \item Tawn type II copula.
$$C(u,v; \delta,\theta)=u^{1-\delta}  \exp\left(-\left[(-\delta\log(u))^\theta+(-\log(v))^\theta\right]^{1/\theta}\right),$$
where $\theta\ge 1$, and  $0\le \delta \le1$ is the skewness parameter. It is known in the literature that the Tawn type II can model the right skewness of copula density.
\item BB8 copula.
$$C(u_1,u_2; \delta,\theta)=1/\delta \left[1-\left(1-\left(1-\bar \delta^{\theta}\right)^{-1}\left(1-(1-\delta u_1)^\theta\right)\left(1-(1-\delta u_2)^\theta\right)\right)^{1/\theta}\right],$$
where $\theta\ge 1$, and  $0\le \bar \delta=1-\delta \le 1$.
\end{itemize}

\subsection{ARMA, GARCH, DCC, and VAR models}
ARMA (Auto Regressive  Moving Average)
and GARCH (Generalized Auto Regressive Conditional Heteroskedasticity) models are  widely-used time series models \cite{CC2008}.  
The ARMA($p$,$q$) model has the general form:
 \[ X_t = \mu + \sum_{k=1}^{p} \phi_k X_{t-k} + \sum_{l=1}^{q}\theta_l\epsilon_{t-l} + \epsilon_t \] where $\phi_k$, $\theta_{l}$ are the parameters of AR and MA, $\mu$ is the intercept, and   $\epsilon_{t}$ is the innovation of the time series. For GARCH$(p,q)$ model, it can be rewritten as
$X_t=\sigma_t \epsilon_t$, and 
$$\sigma_t^2=w+\sum_{j=1}^q \alpha_j \epsilon_{t-j}^2+\sum_{j=1}^p \beta_j\sigma^2_{t-j},$$
where $\alpha_j$ and $\beta_j$ are the coefficients, $\sigma_t^2$ is the conditional variance, and $w$ is the intercept. 

The DCC (Dynamic Conditional Correlation) model introduced in \cite{engle2002dynamic} provides a very good approximation to a variety of time-varying correlation processes.    Let ${\bf x}_t = (x_{1,t},x_{2,t},...,x_{m,t}$) be a vector for $m$-dimensional time series at time $t$. A multivariate GARCH model can be defined as  \[{\bf x}_t = H_t^{1/2}{\bm \epsilon}_t\] where $H_t$ is an $m\times m$ conditional covariance matrix, and ${\bm \epsilon}_t$ is an $m\times 1$ vector of error with mean and variance given by $E({\bm \epsilon}_t) = {\bm 0}$ and ${\rm Var}({\bm\epsilon}_t) = I_m$ where $I_m$ is an $m\times m$ identity matrix. The covariance matrix $H_t$ can be decomposed into 
\[H_t = D_tR_tD_t\]
where $D_t$ = ${\rm diag}({\sqrt{h_{11,t}},...,\sqrt{h_{mm,t}}} )$ is a diagonal of time varying standard deviations from a univariate GARCH$(p,q)$ model, and $R_t$ is a time varying positive definite conditional correlation matrix
\[R_t = {\rm diag}(Q_t)^{-1/2}Q_t{\rm diag}(Q_t)^{-1/2}\]
where \[Q_t = (1-a-b)\Bar{Q} + a{\bf z}_{t-1}{\bf z}'_{t-1} + bQ_{t-1}\] is a positive symmetric matrix, and $\Bar{Q}$ is the unconditional matrix of the standardized errors ${\bf z}_t = D_t^{-1}{\bm\epsilon}_t$. The condition of  $a + b <1$ is imposed to ensure the  stationarity and positive definiteness of $Q_t$. The DCC model consists of two steps: i) The first step is to estimate the univariate GARCH parameters; ii) The second step is to estimate the conditional correlation $R_t$. For more details on the DCC model, please refer to \cite{engle2002dynamic}.  


In the literature of time series, the VAR (Vector Auto Regressive) models are commonly used to investigate the dynamic interactions among multivariate time series \cite{tsay2005analysis}. A VAR($p$) model can be represented as  
$${\bf x}_t = A_1{\bf x}_{t-1} + ... + A_p{\bf x}_{t-p} + {\bm \gamma}_t$$
where $A_i$ is $m\times m$ coefficient matrix for $i = 1,2,..,p$ and ${\bm\gamma}_t$ is a $m$-dimensional error process with zero mean and time-invariant positive definite covariance matrix.


\subsection{Accuracy metrics} 
To evaluate the accuracy of the predictive distribution, we use the following two metrics: i) The first one is the most commonly used MAE (Mean Absolute Error), which can be represented as
$$\mbox{MAE}=\frac{1}{m}\sum_{i=1}^{m} |y_{i}-\hat{y}_{i}|,$$
 where $y_{i}$ represents the observed values and  $\hat{y}_{i}$ represents the predicted values, $i=1,\ldots,m$. ii) The second is the CRPS (Continuous Ranked Probability Score) which is  defined as
$$\mbox{CRPS}(F,s)=\int_{\mathbb{R}}(F(y)-\mathbbm{1}\{s\le y\})^2dy,$$
where $F(\cdot)$ denotes the  CDF  and $\mathbbm{1}\{\cdot\}$ denotes the indicator function. The CRPS measures the difference between the  predicted  CDF  and the empirical CDF of observed values, and is a widely-used accuracy measure dealing with probability forecasts \cite{epstein1969scoring,gneiting2007strictly}. A smaller score indicates a better prediction.  

\section{Copula approach to missing data imputation }\label{sec:missing}
 Let $x_{1,t}$s and $x_{2,t}$s
be the observed time series values of TTN and TTI, respectively, $t=1,\ldots,T$. For modeling purpose, we use the time period from January 20, 2012 to December 31, 2018 as the in-sample data (i.e., 1505 pairs of observations with  596 NAs), while use the data from January 01, 2019 to December 31,  2020 as the out-of-sample data (i.e.,  618 pairs of observations with 282 NAs). Since both TTN and TTI have missing data, we use $x^*_{1,t}$s and $x^*_{2,t}$s to represent the missing observations for TTN and TTI, respectively.  We propose a copula approach to imputing the missing data in Algorithm \ref{alg2} for the in-sample data. In the following, we briefly describe the procedure to impute the missing data for TTN and TTI.
\begin{itemize}
    \item[a)] {\em Model the dependence based on the complete pairs of TTNs and TTIs}. In our study, there are 910 completely observed pairs of TTNs and TTIs among the in-sample data. To model the dependence between TTN and TTI, the empirical marginals are used. We select the best copula structure from various bivariate copula families in the {\rm VineCopula} package by the AIC criterion \cite{schepsmeier2015package}.  It is found that Tawn  type 2 copula is selected for modeling the joint dependence, and the estimated parameters are $\theta=3.93$, $\delta=0.67$, and $\tau=0.54$. In Figure \ref{fig:copulaplot}, we display the normal score plot and the fitted contour plot. We observe that there exists a strong right tail dependence between TTN and TTI which fits the fact that the longer TTI is, the longer TTN.
    \begin{figure}[!hbtp]
\centering
\subfigure[Normal score]{\includegraphics[width=.4\textwidth]{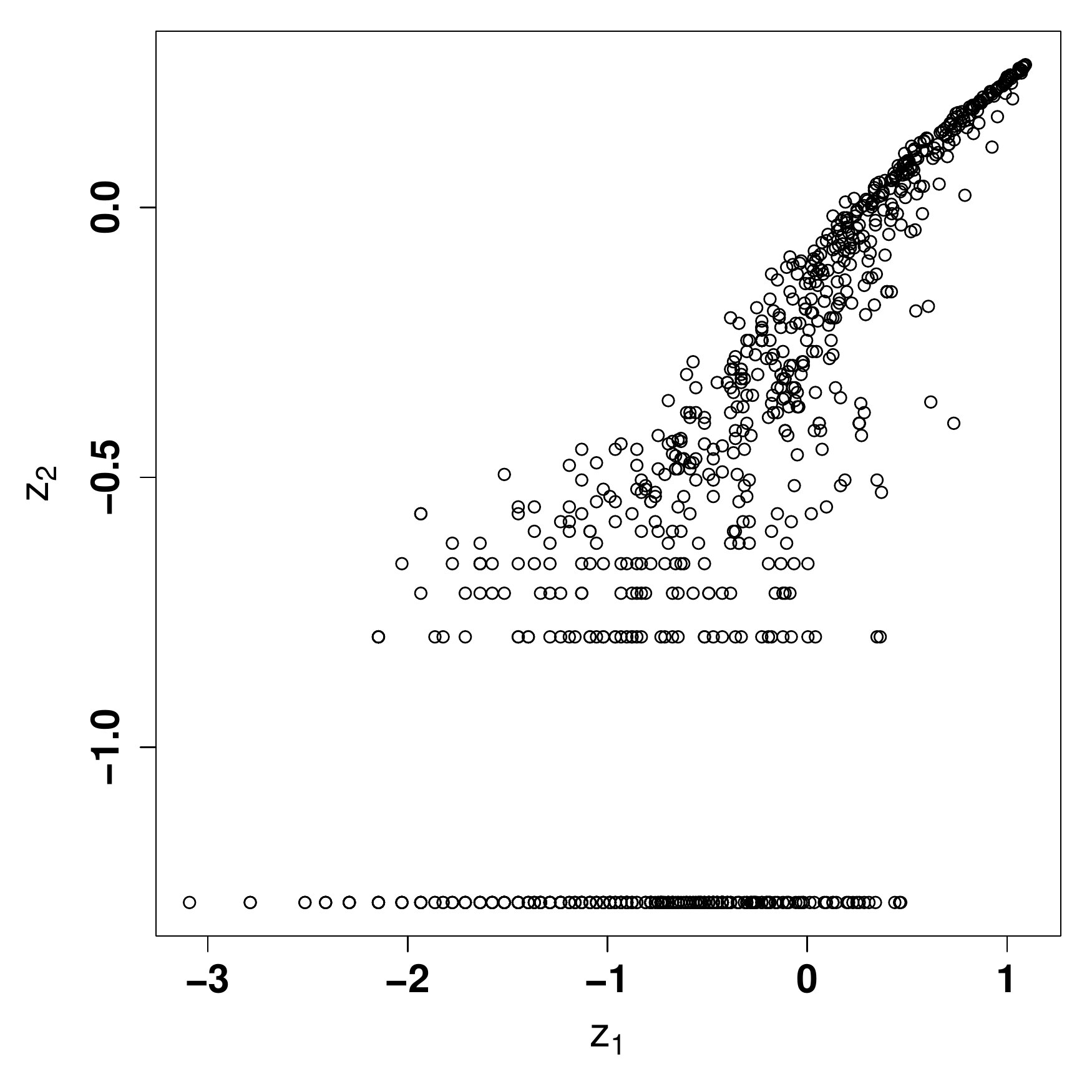} \label{fig:scatter}}
\subfigure[Contour]{\includegraphics[width=.4\textwidth]{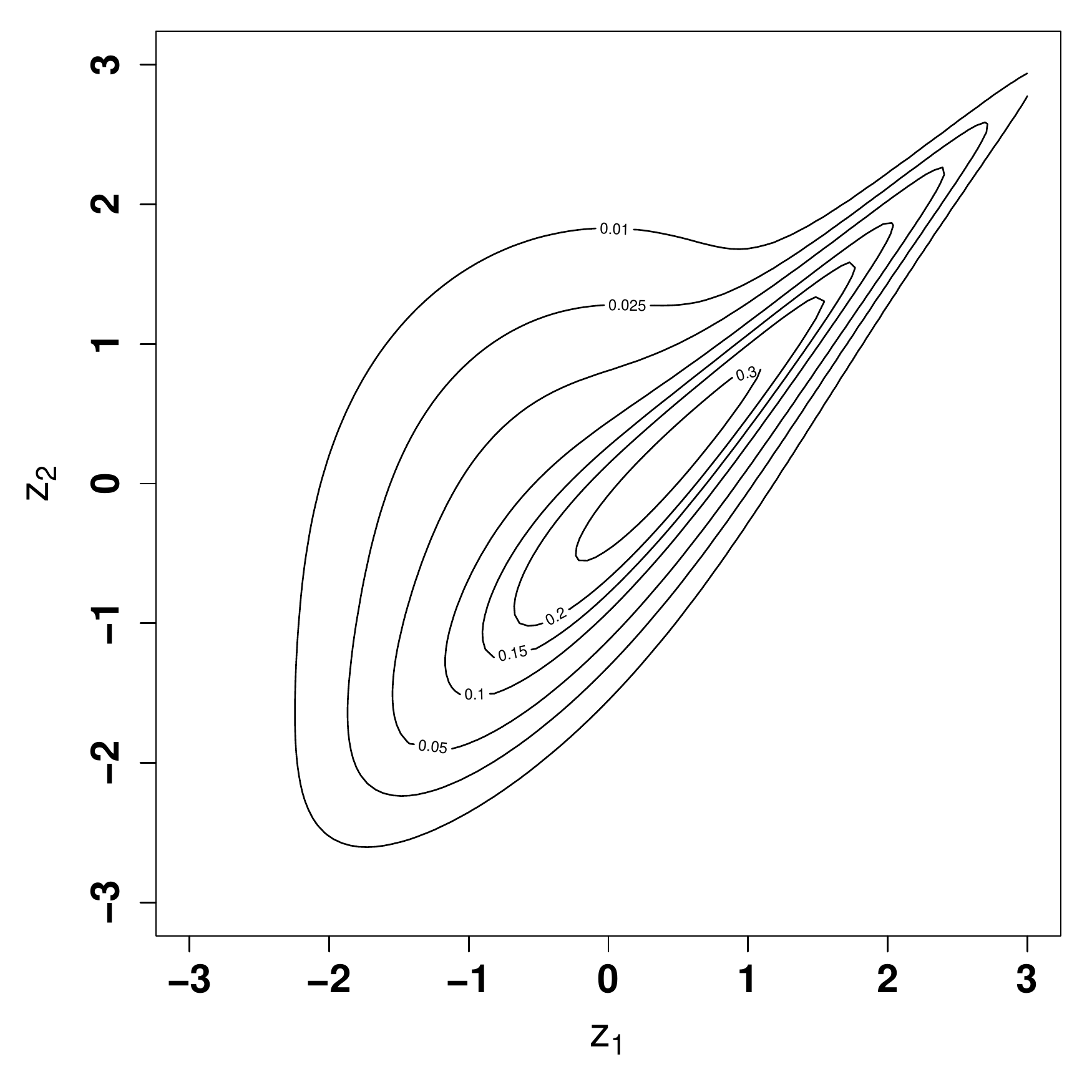} \label{fig:contour}}
\caption{Normal score plot and contour plot of TTI and TTN.  Unit: days. \label{fig:copulaplot}}
\end{figure}
 
    \item[b)] {\em Impute missing data for both TTNs and TTIs.} There are 203 completely missing pairs of TTN and TTI because of unknown breach dates. Based on the estimated copula structure, we simulate $N=5000$ observations in Algorithm \ref{alg2}, where the symbol $^*$  represents the missing value. The missing values are imputed by using the means of simulated observations.

    \item[c)] {\em Impute missing data for TTIs with observed TTNs.} Since there are 392 records with TTNs only because the identification dates are unreported, we impute the missing TTIs by using the conditional copula approach, i.e., sampling from the conditional Tawn  type 2 copula with given TTNs (line 9 in Algorithm  \ref{alg2}). 
\end{itemize}

{  \begin{algorithm}[!hbtp]                      
\caption{Copula approach for imputing the missing data.}          
\label{alg2}                           
\begin{flushleft}
    INPUT:  Complete pairs $\{({x}_{1,t},{ x}_{2,t})|t=i_1,\ldots,i_{T_1}\}$; Incomplete pairs $\{({ x}_{1,t},{x}^*_{2,t})|t=i_{T_1+1},\ldots,i_{T_2}\}$; Missing pairs  $\{({x}^*_{1,t},{ x}^*_{2,t})|t=i_{T_2+1},\ldots,i_{T_3}\}$; $N=5000$.
\begin{algorithmic}[1]                    
\STATE Estimate the empirical marginals $(  u_{1,t}, u_{2,t})$ based on $({x}_{1,t},{ x}_{2,t})$, $t=i_1,\ldots,i_{T_1}$;
\STATE Select the preferred copula structure $C$ via AIC criterion based on the empirical marginals;
    \FOR {$j=1,\ldots,N$}
  \STATE Simulate $T_3-T_2$ pairs from copula $C$;
  \STATE Covert the simulated observations into marginal values $\{({ x}^*_{1,t,j},{ x}^*_{2,t,j})\}$ via the sample quantile approach in \cite{hyndman1996sample}; 
    \ENDFOR
    \STATE ${x}_{1,t}\leftarrow \sum_{j=1}^{N} {x}^*_{1,t,j}/N$, ${x}_{2,t}\leftarrow \sum_{j=1}^{N} {x}^*_{2,t,j}/N$, $t=i_{T_2+1},\ldots,i_{T_3}$;
\FOR {$t=i_{T_1+1},\ldots,i_{T_2}$}
\STATE Simulate $N$ samples of ${u}_{2,t}$'s given each ${u}_{1,t}$ from copula $C$;
\STATE Convert to the original value $x_{2,t}$ based on the mean of ${u}_{2,t}$'s using the sample quantile approach in  \cite{hyndman1996sample};
\ENDFOR
    \RETURN  $\{(t,{ x}_{1,t},{ x}_{2,t})|t=1,\ldots,T_3\}$;
\end{algorithmic}
 OUTPUT: Imputed data $\{(t,{ x}_{1,t},{ x}_{2,t})|t=1,\ldots,T_3\}$ .
\end{flushleft}
\end{algorithm}
}

For comparison purpose, we also use the following two commonly used approaches for missing data imputation. 
\begin{itemize}
    \item  {Kalman Smoothing (KS) imputation.} The KS imputation is the commonly used algorithm and often produces the best performance in the literature  \cite{grewal2014kalman,moritz2017imputets,huyghues2020robust}.
For TTN and TTI, we impute the missing values on the training data by using the KS imputation. 
    Since TTI must be no larger than TTN,  we replace TTI by TTN if the imputed TTI is larger than TTN ({a total of 130 imputed observations}).
    \item  {Multivariate Imputation by Chained Equations (MICE) imputation.}  The second approach that we impose on the missing values is the MICE imputation. This method has emerged as one of the principled method to address missing values in statistical literature \cite{van2011MICE}. Similar to the KS method, we  impute the missing values by the MICE on the training data.  We also replace TTI by TTN when the imputed value of TTI is greater than TTN (a total of 8 imputed observations).
\end{itemize}
\begin{table}[htbp!]\centering
\begin{center}
\begin{tabular}{l|ccccccc}
  \toprule
    &Min &$Q_1$ &Median &Mean &SD &$Q_3$ &Max 
   \\\midrule
 TTN-copula   &0.00  &44.00& 113.0 & 182.7 &251.32 & 190.0 & 2747     \\ \midrule
  TTN-KS   &0.00   &44.00 &106.0 & 180.4&252.08 &208.0&2747   \\ \midrule
   TTN-MICE   &0.00&39.00&87.0&181.2&262.14&213.0  & 2747      \\ \midrule\midrule
TTI-copula   &0.00&1.00&41.00&98.43&200.10&99.19&2196   \\\midrule
TTI-KS   &0.00&3.00 &51.00
&89.46&171.53  &104.00&2196 \\\midrule
TTI-MICE   &0.00&0.00 &32.00&123.5&236.39  &143.0&2196 \\
\bottomrule
\end{tabular}
\end{center}
\caption{Statistics of imputed TTI (time to identification)  and TTN (time to notification) based on different approaches where `SD' stands for standard deviation, $Q_1$, $Q_3$ represent the first and third quantiles.\label{table:summary-imputed}}
\end{table}

Table \ref{table:summary-imputed} shows the summary statistics of imputed TTNs and TTIs for the proposed copula, KS, and MICE approaches. For TTN, the means are very close for different approaches  while the medians are different. The MICE imputation has the smallest median 87 while the copula approach has the largest 113. In terms of standard deviation, the copula approach has the smallest value  251.32  while the MICE approach has the largest value  262.14. For TTI, the means are different, and the smallest mean is 89.46 by the KS imputation while the largest one is 123.5 by the MICE imputation. The MICE approach has the smallest median 32 while the KS approach has the largest 51. For the standard deviation, the KS approach has the smallest value 171.53, however, the MICE approach has the largest value 236.39. It is interesting to note that the copula approach leads to the smallest $Q_3$s for both TTN and TTI.  Compared with KS and MICE imputation approaches, the proposed copula approach is simple but efficient. We further note that all the imputed TTNs are no less than the corresponding TTIs by the copula approach. 

Since both TTNs and TTIs are skewed and have large variabilities, the transformations are performed. For TTNs, since there are only two 0s by the copula imputation, we  replace the 0s by two random values from a uniform $(0,1)$ distribution and perform the log transformation. For TTIs, since we have a  large portion of 0s, we perform the square root transform to reduce the variability.


 \section{Statistical modeling}\label{sec:model}
\begin{figure}[!hbtp]
\centering
\subfigure[TTN]{\includegraphics[width=.4\textwidth]{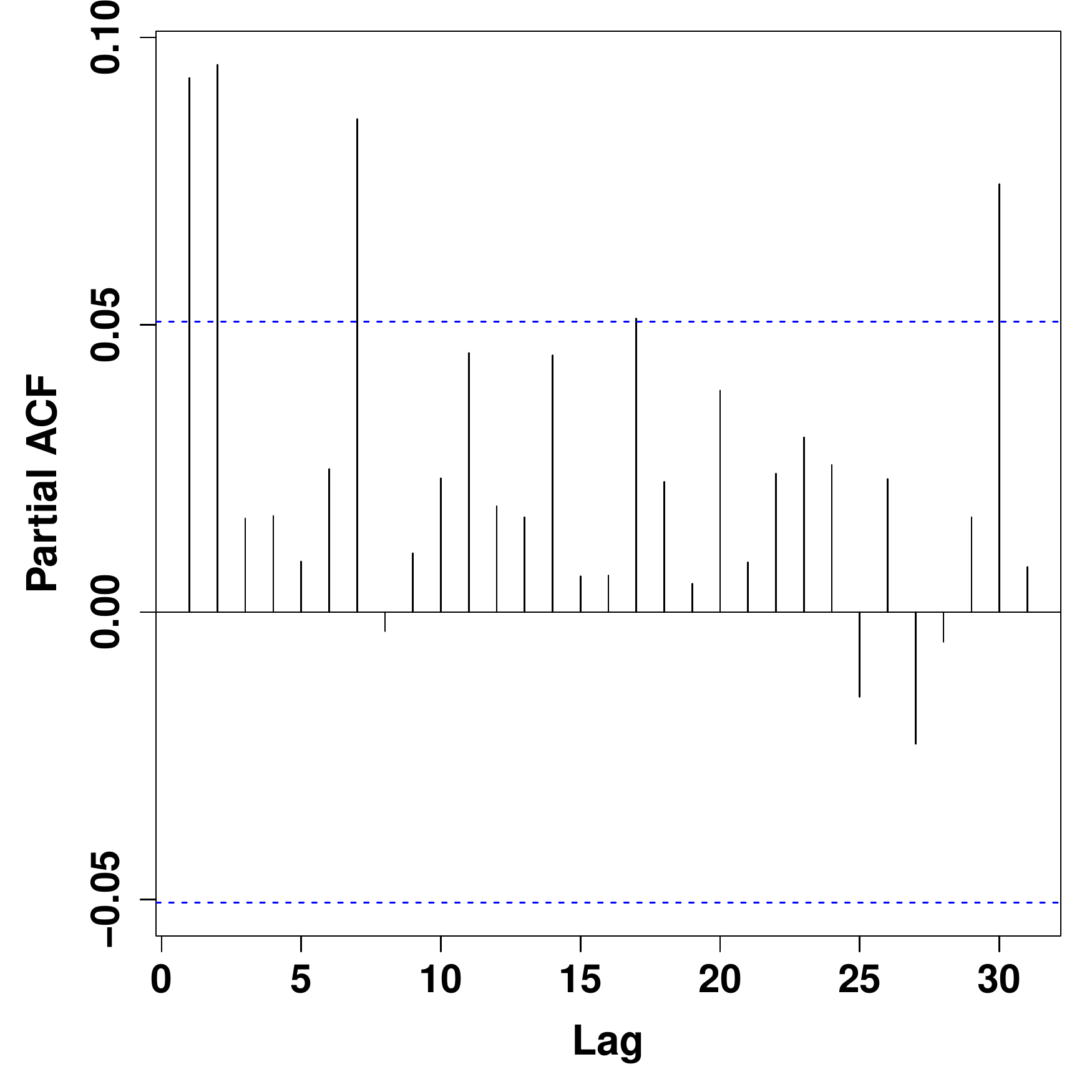} \label{fig:pacfrb}}
\subfigure[TTI]{\includegraphics[width=.4\textwidth]{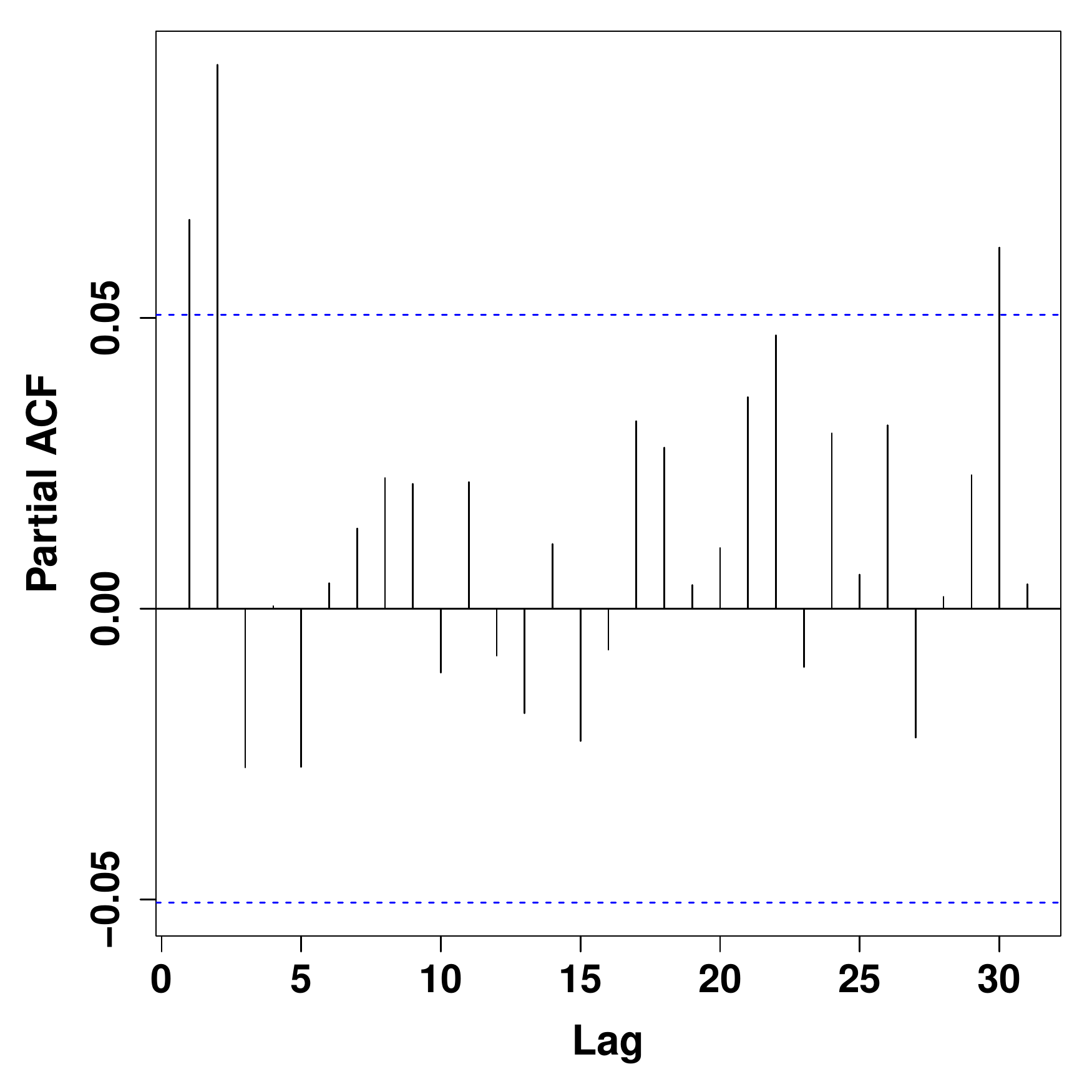} \label{fig:pacfdb}}
\caption{PACFs of transformed TTI and TTN.  Unit: days. \label{fig:pacfplot}}
\end{figure}

 In this section, we develop a copula approach to jointly modeling the dynamics of TTN and TTI. After imputing the missing values, both transformed TTN and TTI exhibit the temporal correlations as shown by their PACFs in Figure \ref{fig:pacfplot}. 

In the following, we discuss how to capture the temporal and cross-sectional dependence of TTN and TTI.
 \subsection{Model fitting}
 From  Section \ref{sec:eda}, it is seen that there exist large variabilities for both TTN and TTI. Therefore, we propose to use a GARCH model to model the volatilities for both TTN and TTI. The analysis on the residuals suggests that GARCH$(1,1)$ is sufficient to describe the volatilities in the residuals of both series. This in fact coincides with the conclusion drawn in the literature that higher-order GARCH models are not necessarily better than GARCH$(1,1)$ \cite{hansen2005forecast}.  Therefore, we fix the GARCH part as GARCH$(1,1)$. To model the evolution of the means of TTN and TTI, we propose using the ARMA$(p,q)$ process. This leads to the following ARMA$(p,q)$+GARCH(1,1) model
\begin{equation}\label{argarch}
X_{t}= \mu+\sum_{k=1}^p \phi_{k}  X_{t-k} +\sum_{l=1}^q \theta_{l}  \epsilon_{t-l} +\epsilon_{t},
\end{equation}
where $\epsilon_{t}=\sigma_{t} Z_{t}$ with $Z_{t}$ being the i.i.d. innovations, {and the $\phi_k$'s and the $\theta_l$'s are respectively the coefficients of the AR and MA parts}. For the standard GARCH$(1,1)$ model, we have
\begin{equation}\label{eq:garch}
\sigma_{t}^2=w+ \alpha_{1} \epsilon^2_{t-1}+  \beta_{1} \sigma^2_{t-1},
\end{equation}
where $\sigma^2_{t}$ is the conditional variance and $w$ is the intercept.
For model selection, we use the AIC criterion to determine the orders of the ARMA models. Note that if ARMA$(p,q)$+GARCH can successfully accommodate the serial correlations in the conditional mean and the conditional variance, there would be no autocorrelations left in the standardized and squared standardized residuals. When the AIC criterion suggests to select multiple models with similar AIC values, we select the simpler model. The autoregressive $p$ and the moving average order $q$ are allowed to vary between $0$ and $5$.  We find that ARMA$(1,1)$+GARCH$(1,1)$ with normal  innovations is sufficient to remove the serial correlations for both TTN and TTI. Based on the Ljung-Box test \cite{brockwell2016introduction}, for TTN, the $p$-values of standardized and squared standardized residuals are $0.186$ and $0.613$ , respectively; for TTI, the $p$-values are $0.524$ and $0.184$, respectively.

Let ${\bf Z}_t=(Z_{1,t},Z_{2,t})$ be the vector of standardized residuals of fitted models for TTN and TTI. Further, we assume that  ${\bf Z}_t=(Z_{1,t},Z_{2,t})$ has the following distribution
\begin{equation}\label{copula-model}
 P\left({\bf Z}_t\le {\bf z}_t\right)=C\left(F(z_{1,t}),G(z_{2,t})\right),
 \end{equation}
where  $F$ is the marginal distribution of $Z_{1,t}$, and $G$ is the marginal distribution of $Z_{2,t}$.
The joint log-likelihood function of the model can be rewritten as
\begin{eqnarray*}
L=
\sum_{t=1}^n \left[\log c\left(F\left(z_{1,t}\right), G\left(z_{2,t}\right)\right)-\log(\sigma_{1,t})-\log(\sigma_{2,t})+ \log\left(f\left( z_{1,t}\right)\right)+ \log\left(g\left(z_{2,t}\right)\right)  \right],
\end{eqnarray*}
where $c(\cdot)$ is the copula density  of $C(\cdot)$; $\sigma_{1,t}$ and $\sigma_{2,t}$ are  the conditional standard deviations of TTN and TTI, respectively.  $f(\cdot)$ is the density function of $Z_{1,t}$, and $g(\cdot)$ is the density function of $Z_{2,t}$. A popular method for estimating the parameters of a joint model is the Inference Function of Margins method \cite{Joe1997}, which is employed in our study. This method has two steps: (i) estimate the parameters of the marginal  models; and (ii) estimate the parameters of the copula by fixing the parameters obtained at step (i). Since we have identified the time series models for TTN  and TTI, we discuss how to model the bivariate dependence in the following. {Note that although we assume the normal innovations for the marginal processes to remove serial correlations, $Z_{1,t}$s and $Z_{2,t}$s are not normally distributed  due to the high skewness and an excessive number of 0s}. Since it is very challenging to fit parametric distributions to the marginals, we use the empirical marginals in Eq. \eqref{copula-model}, and then select the copula structure by using the AIC criterion.  The BB8 copula is selected to model the dependence between the standardized residuals, and the corresponding estimated parameters are $\theta=4.54$, $\delta=0.98$, and $\tau=0.64$.

\begin{figure}[!hbtp]
\centering
\subfigure[Normal score]{\includegraphics[width=.4\textwidth]{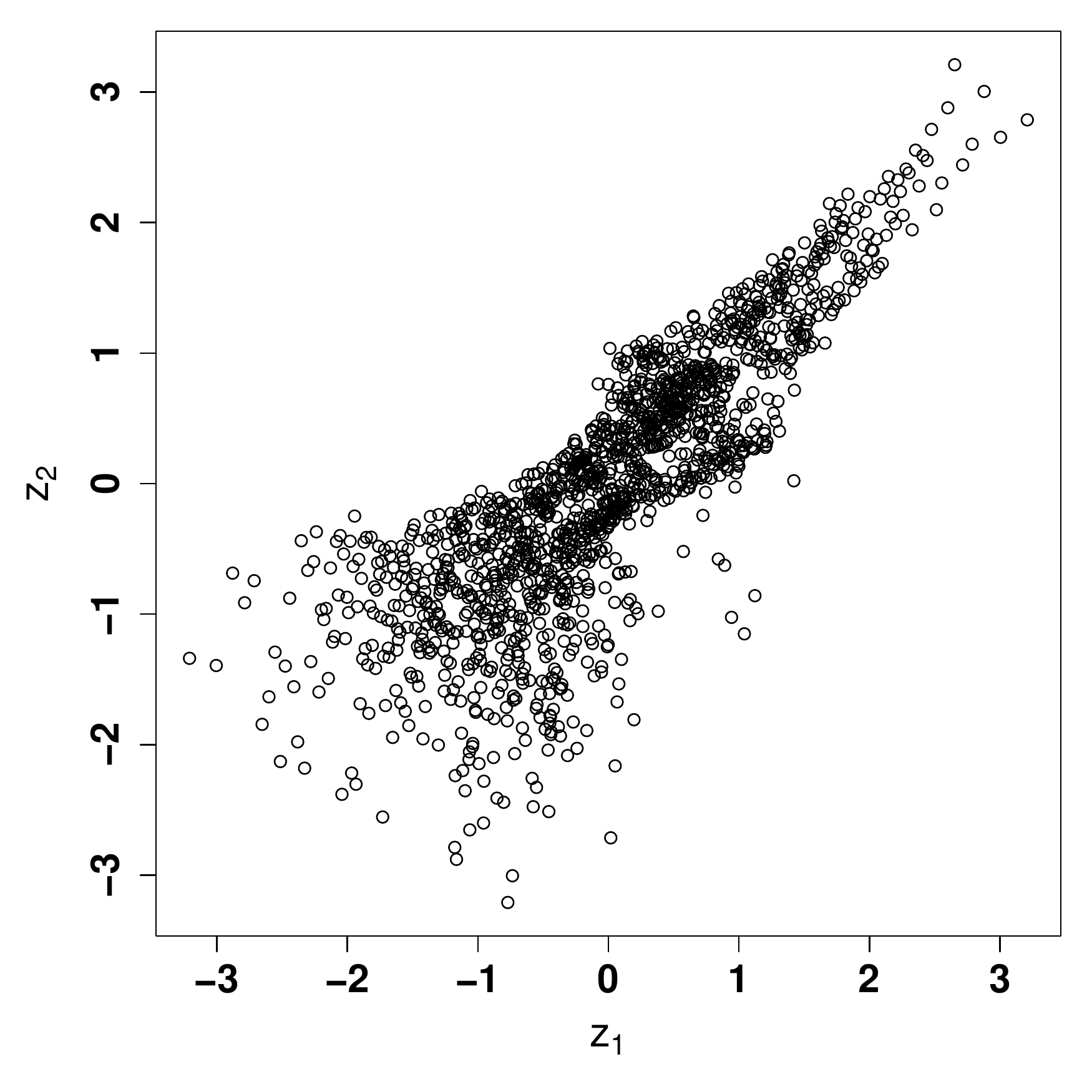} \label{fig:scatter2}}
\subfigure[Contour]{\includegraphics[width=.4\textwidth]{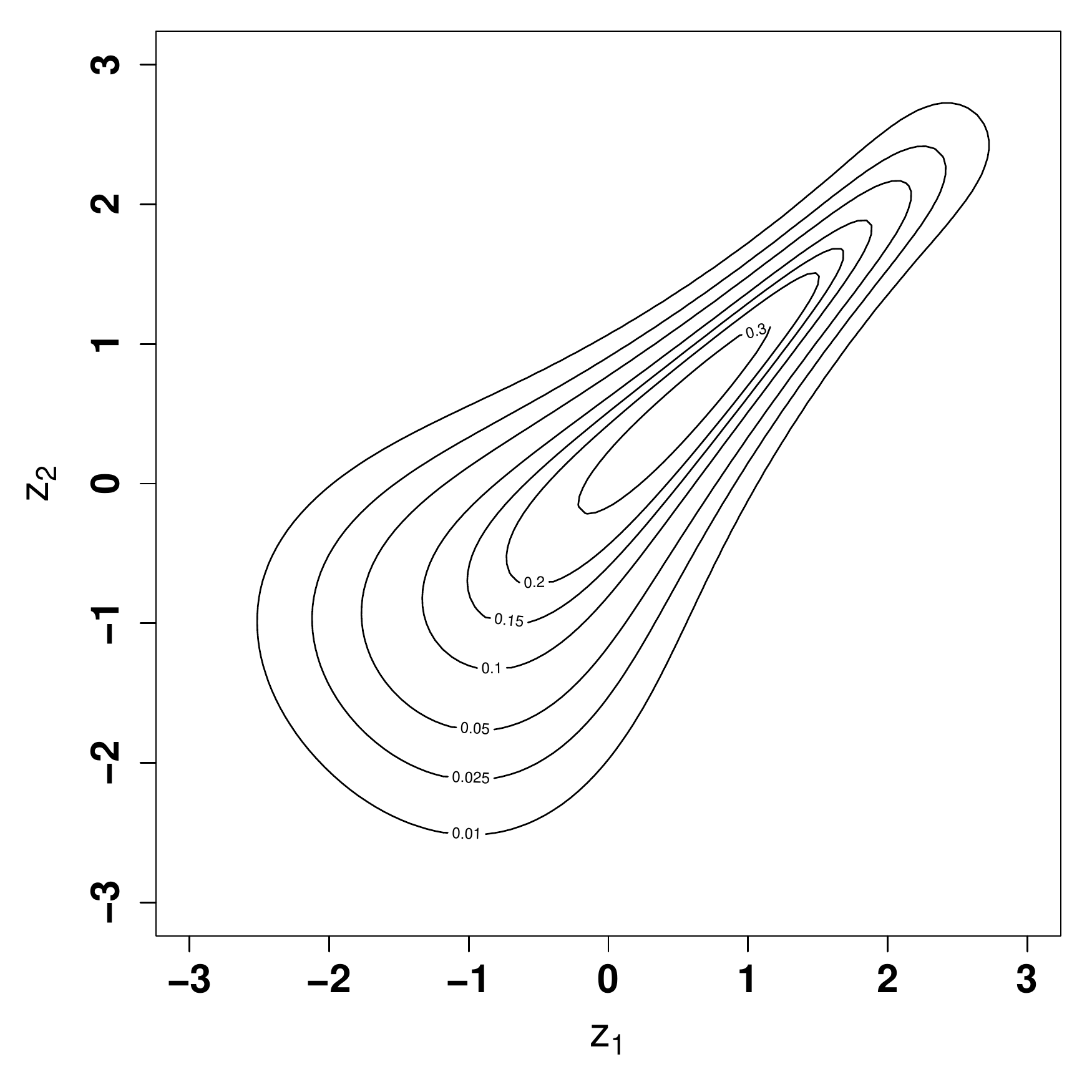} \label{fig:contour2}}
\caption{Normal score plot and contour plot of residuals of TTI and TTN. \label{fig:copulaplot1}}
\end{figure}
The normal score plot and fitted contour plot are displayed in Figure \ref{fig:copulaplot1}. It is seen that the upper tail dependence is  well captured by the BB8 copula.

  \subsection{Prediction evaluation}
We use Algorithm \ref{alg:var1} to perform the rolling window prediction for TTI and TTN. The parsimonious ARMA(1,1)+GARCH(1,1) model is applied to the sample data with window size $l=500$. The window size is selected based on the fact that  too few observations can lead to the large variability in the model and hence have a poor predictive performance, while too many observations not only require more computational effort but also not necessarily improve the predictive performance because of the potential structure break and trends. In the rolling process, the dependence structure is allowed to vary with time. That is, the copula is re-selected during the fitting process via the criterion of AIC (see line 4 of  Algorithm \ref{alg:var1}).  Since the size of out-of-sample data is 618, we have $T=1505$ and $S=2123$. The predictive distributions of TTN and TTI are simulated based on $N=5000$ samples. If the observed value is missing, we use the predicted mean to replace the missing value in the TTN/TTI to perform the rolling prediction. The evaluation metrics such as MAE and CRPS are computed by excluding the missing data in the out-of-sample data.
\begin{algorithm}[!hbtp]
{ 
\caption{{ Algorithm for predicting the distributions of TTN  and TTI.}\label{alg:var1}}
\begin{flushleft}
Input: Imputed in-sample data $\{(t,x_{1,t},x_{2,t})|t=1,\ldots,T\}$; out-of-sample data $\{(t,x_{1,t} ,x_{2,t})|t=T+1,\ldots,S\}$; window size $l=500$; $N=5000$.
\begin{algorithmic}[1]
\FOR {$i = T,\cdots,S-1$}
\STATE{Estimate the ARMA(1,1)+GARCH(1,1) model based on the log-transformed TTNs with time window $[i-l+1,i]$;}
\STATE{Estimate the ARMA(1,1)+GARCH(1,1) model based on the square root transformed TTIs with time window $[i-l+1,i]$;}
\STATE{Convert the standardized residuals $z_{1,t}$s and  $z_{2,t}$s to the empirical marginals;}
\STATE{Select a preferred copula based on the empirical marginals via the AIC criterion;}
\STATE{Based on the estimated copula, simulate $N$ $2$-dimensional copula samples $\left(u^{(k)}_{1,i},u^{(k)}_{2,i}\right)$, $k=1,\ldots,N$;}
\STATE{For the TTN, convert the simulated dependent samples $u^{(k)}_{1,i}$s into standardized residuals $z^{(k)}_{1,i}$s   via the sample quantile approach in \cite{hyndman1996sample},
$k=1,\ldots,N$;}
\STATE{For the TTI, convert the simulated dependent samples $u^{(k)}_{2,i}$s into standardized residual $z^{(k)}_{2,i}$s  via the sample quantile approach in \cite{hyndman1996sample},
$k=1,\ldots,N$;}
\STATE{Compute the predicted $N$  $2$-dimensional $x^{(k)}_{1,i+1}$s and $x^{(k)}_{2,i+1}$s, respectively, $k=1,\ldots,N$;}
\IF{$x_{1,i+1}$ ($x_{2,i+1}$) is missing}
\STATE{$x_{1,i+1}$ ($x_{2,i+1}$) is imputed as the mean of $x^{(k)}_{1,i+1}$s ($x^{(k)}_{2,i+1}$s);}
\ENDIF
\ENDFOR \\
\end{algorithmic}
Output: Predictive distributions of TTN and TTI.
\end{flushleft}
}
\end{algorithm}
\paragraph{Imputation comparison.} We   compare the predictive performance of proposed copula imputation approach to those of KS and MICE imputation approaches. For this purpose, Algorithm \ref{alg:var1} is also employed to the KS and MICE imputed data. The predictive results are reported in Table \ref{table:copula-imp}.  It is seen that for TTN,  the predictive performances are similar for all three imputation approaches in terms of the mean of CRPSs and MAE. For TTI, the copula imputation approach leads to a slightly smaller mean of CRPSs while the KS imputation approach has a slightly smaller MAE. We also compute the percentages that the
CRPSs of the copula approach are less than that of MICE and KS, respectively. We observe that  the copula approach outperforms both of the MICE and KS imputation approaches in terms of CRPSs. In particular, for TTI, the copula approach improves the MICE approach by 22.71\% and KS approach by 11.93\%.
 
  \begin{table}[htbp!]
\begin{center}
\begin{tabular}{l|c|c||c|c}
\hline
    &CRPS-Mean   &MAE     &CRPS-Mean   &MAE    \\
\hline 
&\multicolumn{2}{c||}{TTN}&\multicolumn{2}{c}{TTI}  \\ \hline
{\bf Copula }&  {100.555}   & 141.230  &84.896& 130.308       \\
\hline
MICE &100.581    &141.359&85.300 & 134.282  \\
\hline
KS & 100.478  &141.493& 85.998  &129.997            \\
\hline
&\multicolumn{4}{c}{Percentage}\\\hline 
Copula vs MICE    & \multicolumn{2}{c||}{52.24\%}    &\multicolumn{2}{c}{72.71\%}    \\ \hline
Copula vs KS    & \multicolumn{2}{c||}{55.86\%} &\multicolumn{2}{c}{61.93\%}   \\
\hline
\end{tabular} 
\end{center}
\caption{Means of CRPSs and MAEs of the proposed copula  and other  imputation approaches. The {\em Percentage} is the \% of the CRPSs of the proposed imputation approach that are less than that of the other approaches.  \label{table:copula-imp}}
\end{table}
Therefore, the proposed copula imputation approach is preferred and used in the following discussion.

\paragraph{Model comparison.} 
We compare the predictive performance of proposed model to those commonly used models of DCC and VAR. For   fair comparison,  modified Algorithms \ref{alg:var1}  are employed to the DCC and VAR models. Specifically, for the DCC model, the marginals of TTN and TTI are still fitted by using ARMA(1,1)+GARCH(1,1) on the sliding window, but the DCC is used for modeling the correlation. We also simulate $N=5000$  values for each prediction for the DCC model. For the VAR model, the order $p$ is selected from $1$ to $5$ by using the AIC criterion for each sliding window, and $N=5000$ predicted values are simulated from the selected VAR model for each prediction.    
  \begin{table}[htbp!]
\begin{center}
\begin{tabular}{l|c|c||c|c}
\hline
    &CRPS-Mean   &MAE     &CRPS-Mean   &MAE    \\
\hline 
&\multicolumn{2}{c||}{TTN}&\multicolumn{2}{c}{TTI}  \\ \hline
  {\bf Copula }  & 100.555 & 141.230     & 84.896& 130.308    \\ \hline
  DCC  & 101.022 & 144.146     & 86.422&129.550     \\
\hline
VAR &101.233   &143.234  & 108.532 &100.465\\
\hline
&\multicolumn{4}{c}{Percentage}\\\hline 
Copula vs DCC   & \multicolumn{2}{c||}{45.69\%}   & \multicolumn{2}{c}{72.94\%} \\ \hline
Copula vs VAR    & \multicolumn{2}{c||}{57.24\%}     & \multicolumn{2}{c}{71.10\%}  
\\
\hline
\end{tabular} 
\end{center}
\caption{Mean CRPSs and MAEs of the proposed copula  and other model(s). The {\em Percentage} is the \% of the CRPSs of the proposed model that are less than that of the other model(s).   \label{table:rps-models}}
\end{table}

The predictive results are reported in Table \ref{table:rps-models}. It is seen that for TTN, the predictive performances are comparable based on MAE and mean CRPS. However, in terms of individual CRPS, DCC is slighter better than the proposed copula approach as it improves 4.31\%. For TTI, it is seen that the VAR model has the smallest MAE. But the proposed copula approach significantly outperforms other approaches in terms of individual CRPS, namely improving 22.94\% compared to DCC and 21.10\% compared to VAR.

To further assess the prediction accuracy,  we further use the Value-at-Risk (VaR) \cite{mcneil2015quantitative} metric since it is directly related to the high quantiles of interest.  Recall that for a random variable $X_t$, the VaR at level $\alpha$ for some $0<\alpha<1$ is defined as
${\rm VaR}_\alpha(t)=\inf\left\{l: P\left(X_t \leq l\right)\ge \alpha\right\}$.
For example, ${\rm VaR}_{.95}(t)$ means that there is only a $5\%$ probability that the observed value is greater than the predicted value ${\rm VaR}_{.95}(t)$.
An observed value that is greater than the predicted ${\rm VaR}_\alpha(t)$ is called a {\em violation}. In order to evaluate the prediction accuracy of the VaR values, we use the following three popular tests \cite{christoffersen1998evaluating}:
(i) the unconditional coverage test, denoted by LRuc, which evaluates whether or not the fraction of violations is significantly different from the model's violations; (ii) the conditional coverage test, denoted by LRcc, which is a joint likelihood ratio test for the independence of violations and unconditional coverage; and (iii) the dynamic quantile test (DQ), which is based on the sequence of `hit' variables \cite{engle2004caviar}.
  \begin{table}[htbp!]
\begin{center}
\begin{tabular}{l|c|c|c|c|c||c|c|c|c|c}
\hline
    &Exp.  &Ob.     &LRuc   &LRcc &DQ  &Exp.  &Ob.    &LRuc   &LRcc &DQ \\  \hline
 &\multicolumn{5}{c|}{TTN}&\multicolumn{5}{c}{TTI}  \\ \hline
&   \multicolumn{10}{c}{${\rm VaR}_{.9}$}\\
\hline 
{\bf Copula }&   58  &  54 & .576  &   .484  &.999   &43.6 &46 &  .704 & .795   & .999   \\
\hline
VAR &58  & 39& .005 &  .019 & .410  &43.6&147    &0&0&0 \\
\hline
DCC & 58  & 39  & .005    & .019  & .394  &43.6&31 &    .035 &   .107   & .758    \\
\hline
&   \multicolumn{10}{c}{${\rm VaR}_{.95}$}\\
\hline 
 {\bf Copula }&  29   & 26 & .561   &   .249   &  .995   &21.8 &22 & .965&  .993   & .999    \\
\hline
VAR &29   & 22  & .164  &  .159 & .952&21.8&139 &0&0&0 \\
\hline
DCC &29  &  21 &  .110   & .126  &  .891&21.8 &   21    & .860  &  .984   &  .999        \\\hline
\hline
&   \multicolumn{10}{c}{${\rm VaR}_{.99}$}\\
\hline 
 {\bf Copula }&  5.8 &7 &.001& .003& .568   &4.36 &    5 &.003& .013&.743  \\
\hline
VAR &5.8&7&.001&.003&.566&4.36& 127 &0&0&0   \\
\hline
DCC &5.8   & 11&.001&.003&.570&4.36    &13  & 0&0&0      \\\hline
\end{tabular} 
\end{center}
\caption{The $p$-values of the VaR tests of the predicted violations for $\alpha=.9,.95,.99$.   `Ob.' represents the observed number of violations and
 `Exp.' represents the expected number of violations.  \label{table:vio}}
\end{table}

Table \ref{table:vio} shows the $p$-values of VaR tests at different levels of $\alpha=.9, .95, .99$. It is seen that at $\alpha=.9$ and $\alpha=.95$  levels, the copula approach predicts very well. For  TTN, we observe that the numbers of expected violations are very close to the numbers of observed violations. The $p$-values are all very large for those three tests. Compared with the proposed copula approach, the discrepancies between the numbers of
expected violations and the number of observed violations are large for the VAR and the DCC models. At level $\alpha=.9$, the $p$-values of LRuc and LRcc are  small  for the VAR and the DCC models. Similarly, for  TTI, the proposed copula approach significantly outperforms the other approaches. Particularly, we observe that the VAR has the worst predictive performance.  At  $\alpha=.99$ level, we observe that the numbers of
expected violation  and the numbers of observed violation are close based on the copula approach for both TTN and TTI. However, the $p$-values are small for both LRuc and LRcc because of the small sample size. But the proposed copula approach still outperforms the other approaches. Figure \ref{fig:copulaplot} shows the  VaR plots of TTN and TTI, and we observe that the proposed copula approach predicts the VaRs well.

   \begin{figure}[!hbtp]
\centering
\subfigure[TTN]{\includegraphics[width=.45\textwidth]{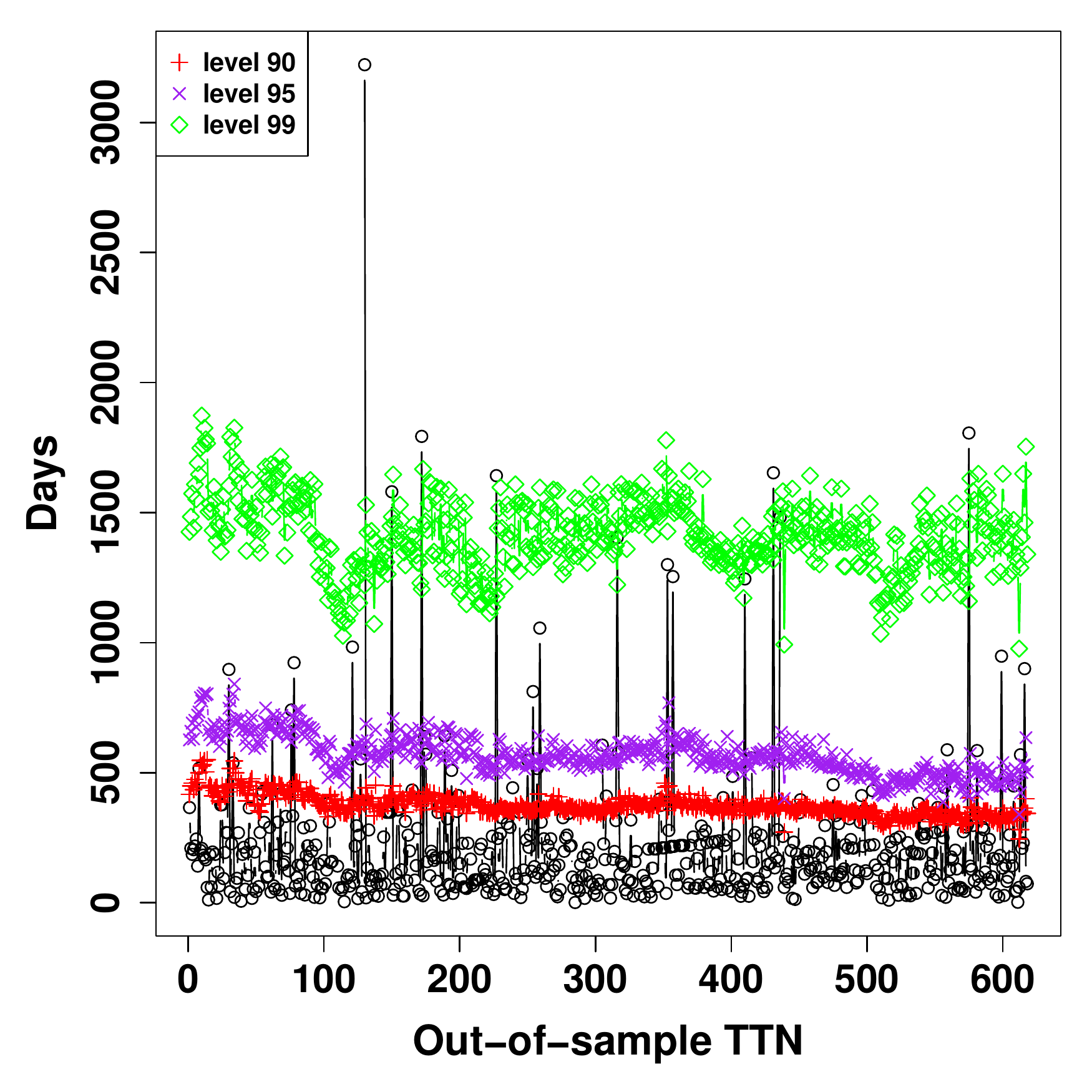} \label{fig:scatter}}
\subfigure[TTI]{\includegraphics[width=.45\textwidth]{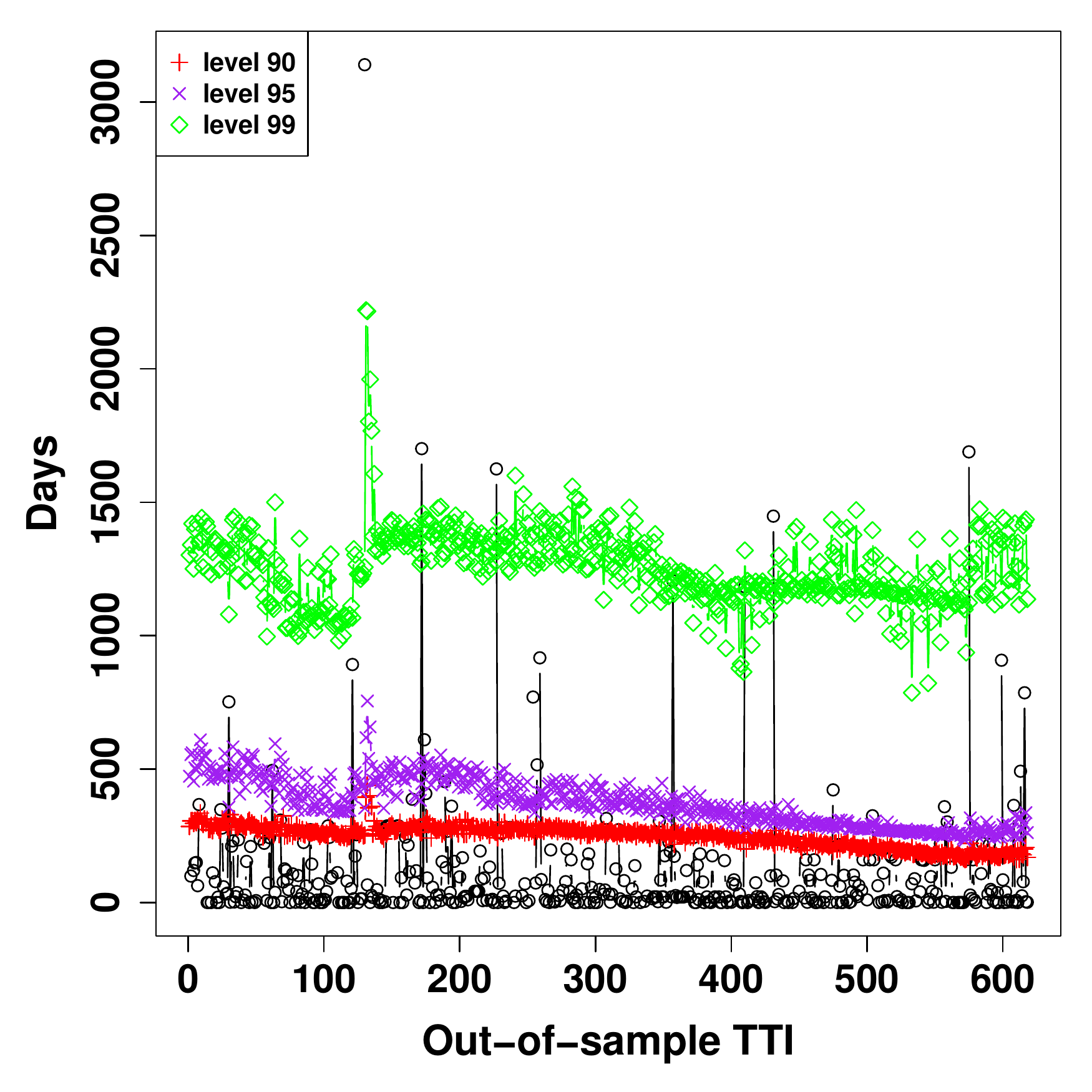} \label{fig:contour}}
\caption{VaR plots of TTN and TTI. \label{fig:copulaplot}}
\end{figure}

To summarize, the proposed copula model outperforms the DCC and VAR models in terms of CRPS and VaR overall.
 



\section{Conclusion and discussion}\label{sec:discussion}
{In this paper, we develop a statistical model for capturing the dependence between two important metrics related to the data breach risk, i.e., TTN and TTI. To tackle the missing data, we propose a novel copula imputation approach. Our study shows that the proposed imputation approach is superior to the other commonly used imputation approaches. We further develop a copula model to capture the dynamics of TTN and TTI. The new model has a satisfactory prediction performance and outperforms the other multivariate time series models such as DCC and VAR. 

From our study, it is discovered that both TTN and TTI have large variabilities. Therefore, insurance companies should not adopt the mean of TTN or TTI as the risk measure although this information is available from the public source (e.g. 2021 Cost of a Data Breach Report by IBM). This can severely underestimate  the relevant cost. We recommend using the VaR as the measurement to reflect the risk. Taking the VaR$_{.95}$ as a representative example, according to the proposed copula model, 95\% of TTNs are less than 619.5  days with a standard deviation 70.769 days in 2019, and it reduces to  534.9  days with a standard  deviation  57.06 days in 2020. For TTI, 95\% of TTIs are less than 456.9 days with a standard  deviation  64.3222 days in 2019, and it reduces to  323.7  days with a standard  deviation  55.307 days in 2020. Therefore, the mean VaR is more suitable for measuring the risk. We also find that there exists a large variability in the differences of TTNs and TTIs. We urge  the authority in CA to require that the notification must be made within a certain period (e.g., 30/60 days). This can efficiently reduce the unnecessary delay from the identification to the notification. 
}

Although the proposed model is based on the CA breach data, the developed approach can be equally applied to the similar breach data. Further, the proposed model can help  insurers to estimate  TTI  and TTN. In the following, we briefly mention how to use the developed model in the risk assessment from an insurer's perspective: 

\begin{itemize}
    \item Pricing factor. Assume that an insurance company offers a cyber insurance policy covering the cost  related to TTN/TTI (e.g., notification expense, regulatory fines and penalties, and forensic expenses, etc). For the pricing formula, the TTN/TTI should be taken into account since it is directly related to the loss. The proposed model can be used to predict the quantities of interest such as  high quantiles of TTN/TTI. Those quantities can be used as a pricing factor to adjust the pricing formula. 
   
    \item Individual incident. It is common in practice that when a cyber incident is identified, the breach date is unknown or needs much time or effort to investigate. The proposed model can be used to estimate/predict the missing/unknown TTN/TTI for the cost estimation.    
\end{itemize}
The current study also has some limitations like other studies. First, the proposed approach is based on the CA breach reports, it is possible that the data from other states or countries may exhibit different patterns. Therefore, the current model should be cautiously used when the different pattern appears. Second, the covariates are not incorporated in our modeling process. In the future study, the text mining approach can be used to extract the key information as the covariates. Third, the severity related to the TTN/TTI is of interest. However, due to the limited loss data, this study will be pursued when more data is available.

\bibliographystyle{plain}  
\bibliography{letter,breach,att-all}

\end{document}